\documentclass[a4paper,fleqn,usenatbib,useAMS]{mnras}

\usepackage{graphicx}	
\usepackage{amsmath}	
\usepackage{amssymb}	
\usepackage{multicol}        
\usepackage{bm}		
\usepackage{pdflscape}	

\usepackage{newtxtext,newtxmath}
\usepackage[T1]{fontenc}
\usepackage[flushleft]{threeparttable}
\usepackage[draft]{changes}

\newcommand{\gsim}{\;\lower.6ex\hbox{$\sim$}\kern-7.75pt\raise.65ex\hbox{$>$}\;}
\newcommand{\lsim}{\;\lower.6ex\hbox{$\sim$}\kern-7.75pt\raise.65ex\hbox{$<$}\;}

\newcommand{\MSUN}{$M_{\odot}$}

\title[]{The spatially resolved star formation history of the dwarf spiral galaxy NGC 5474}

\author[G. Bortolini et al.]{
	G. Bortolini$^{1,}$$^{2}$\thanks{E-mail: giacomo.bortolini@astro.su.se}, 
	M. Cignoni$^{2,}$$^{3,}$$^{5}$\thanks{E-mail: michele.cignoni@unipi.it},
    E. Sacchi$^{4,}$$^{5}$,
    M. Tosi$^{5}$,
    F. Annibali$^{5}$,
    R. Pascale$^{5}$,
    M. Bellazzini$^{5}$,
    \newauthor
    D. Calzetti$^{6}$,
    A. Adamo$^{1}$,
    Daniel. A. Dale$^{7}$,
    M. Fumagalli$^{8}$,
    John. S. Gallagher$^{9}$,
    K. Grasha$^{10}$\thanks{ARC DECRA Fellow},
    \newauthor
    Kelsey E. Johnson$^{11}$,
    Sean. T. Linden$^{6}$,
    M. Messa$^{5}$,
    G. Östlin$^{1}$,
    E. Sabbi$^{12}$,
    A. Wofford$^{13}$
	\\
	$^{1}$Department of Astronomy, The Oskar Klein Centre, Stockholm University, AlbaNova, SE-10691 Stockholm, Sweden\\
	$^{2}$Department of Physics - University of Pisa, Largo B. Pontecorvo 3, 56127, Pisa, Italy\\
    $^{3}$INFN, Largo B. Pontecorvo 3, 56127, Pisa, Italy\\
    $^{4}$ Leibniz-Institut fur Astrophysik Potsdam, An der Sternwarte 16, D-14482 Potsdam, Germany\\
    $^{5}$INAF – Osservatorio di Astrofisica e Scienza dello Spazio di Bologna, Via Gobetti 93/3, I-40129 Bologna, Italy\\
    $^{6}$Department of Astronomy, University of Massachusetts Amherst, 710 North Pleasant Street, Amherst, MA 01003, USA\\
    $^{7}$Department of Physics and Astronomy, University of Wyoming, 1000 E. University, Laramie, WY 82071, USA\\
    $^{8}$Dipartimento di Fisica G. Occhialini, Universit`a degli Studi di Milano Bicocca, Piazza della Scienza 3, 20126 Milano, Italy\\
    $^{9}$Department of Astronomy, University of Wisconsin–Madison, 475 N. Charter Street, Madison, WI 53706, USA\\
    $^{10}$Research School of Astronomy and Astrophysics, Australian National University, Camberra, Australia\\
    $^{11}$Department of Astronomy, University of Virginia, Charlottesville, VA, USA\\
    $^{12}$Space Telescope Science Institute, 3700 San Martin Drive, Baltimore, MD 21218, USA\\
    $^{13}$Universidad Nacional Autónoma de México Instituto de Astronomía, AP 106 Ensenada, 22860, BC, México\\
}

\date{Accepted XXX. Received YYY; in original form ZZZ}

\pubyear{2023}

\begin{document}
	\label{firstpage}
	\pagerange{\pageref{firstpage}--\pageref{lastpage}}
	\maketitle
	
	\begin{abstract}
		We study the resolved stellar populations and derive the star formation history of NGC 5474, a peculiar star-forming dwarf galaxy at a distance of $\sim 7$ Mpc, using Hubble Space Telescope Advanced Camera for Surveys data from the Legacy Extragalactic UV Survey (LEGUS) program. We apply an improved colour-magnitude diagram fitting technique based on the code SFERA and use the latest PARSEC-COLIBRI stellar models. Our results are the following. The off-centre bulge-like structure, suggested to constitute the bulge of the galaxy, is dominated by star formation (SF) activity initiated $14$ Gyr ago and lasted at least up to $1$ Gyr ago. Nevertheless, this component shows clear evidence of prolonged SF activity (lasting until $\sim 10$ Myr ago). We estimate the total stellar mass of the bulge-like structure to be $(5.0 \pm 0.3) \times 10^{8}$ \MSUN. Such a mass is consistent with published suggestions that this structure is in fact an independent system orbiting around and not within NGC 5474's disc. The stellar over-density located to the South-West of the bulge-like structure shows a significant SF event older than $1$ Gyr, while it is characterised by two recent peaks of SF, around $\sim10$ and $\sim100$ Myr ago. In the last Gyr, the behavior of the stellar disc is consistent with what is known in the literature as `gasping'. The synchronised burst at $10-35$ Myr in all components might hint to the recent gravitational interaction between the stellar bulge-like structure and the disc of NGC 5474.
	\end{abstract}
	


\section{Introduction} \label{intro}

Dwarf galaxies are the most common type of galaxy in the Universe \citep{Mateo1998,Tolstoy2009,Annibali2022}. They display a rich variety of properties (e.g., mass, luminosity, surface brightness, metallicity), and occupy a diverse set of environments \citep[see e.g.,][]{Mateo1998,Tolstoy2009,McQuinn2011,Weisz2011,Weisz2014,Gallart2015,Cignoni2019,Annibali2022,Dale2023}, indicating highly complex and diverse formation scenarios. In the early universe, small dark matter halos are believed to have been the sites of the earliest star formation \citep[see][]{White1991}, and the building blocks of more massive systems \citep{White1978,Peebles1982,Frenk1988}. Therefore, present-day dwarf galaxies play an important role in our understanding of the physical processes related to galaxy formation and evolution (e.g., star formation, stellar feedback, ram pressure, and gravitational interactions).\par 

One of the crucial ingredients to understand how galaxies form and evolve is their  \textit{star formation history} (SFH), i.e. the rate at which gas is converted into stars as a function of age. Observations of high redshift galaxies provide snapshots of the Universe when it was younger than today. However, due to the large distances of these galaxies from us, we cannot study in detail their evolution over the many billions of years since their formation. Thanks to modern space telescopes' spatial resolution and sensitivity capabilities, we can resolve single stars in galaxies within $\sim 20$ Mpc \citep{Tolstoy2009,Cignoni2010,Annibali2022}, and draw their colour-magnitude diagrams (CMDs). Modeling the observed CMD of these galaxies, using state-of-the-art stellar evolution models, is the most reliable method we have to derive stellar ages, and thus recover the SFH over the whole lifetime of the galaxy \citep{Tosi1991,Dolphin2002,Gallart2005,Cignoni2010}.\par

\begin{figure*}
	\centering
	\includegraphics[width=0.45\textwidth]{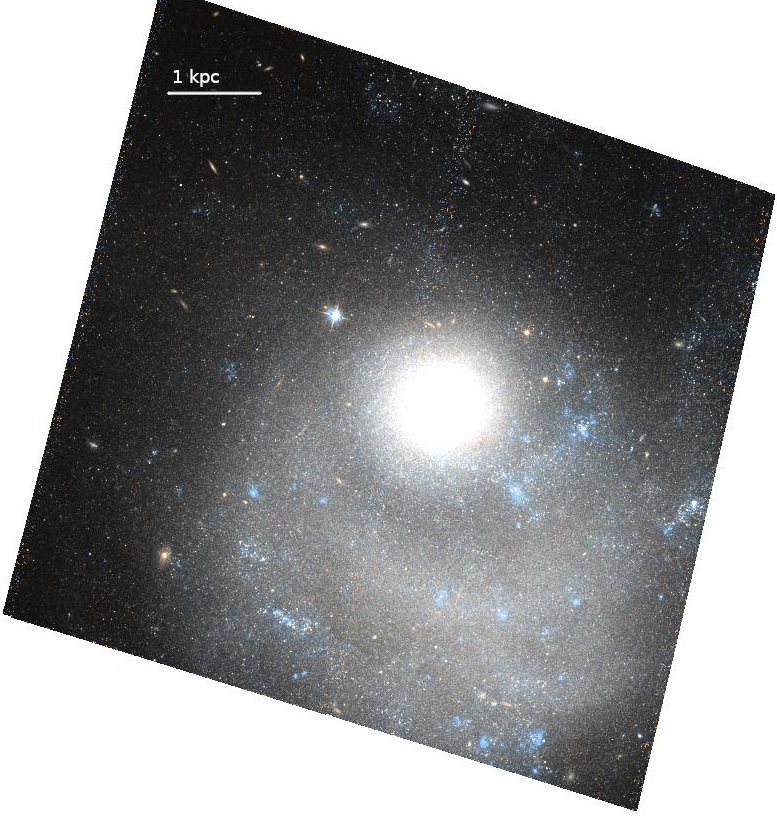}
    \includegraphics[width=0.45\textwidth]{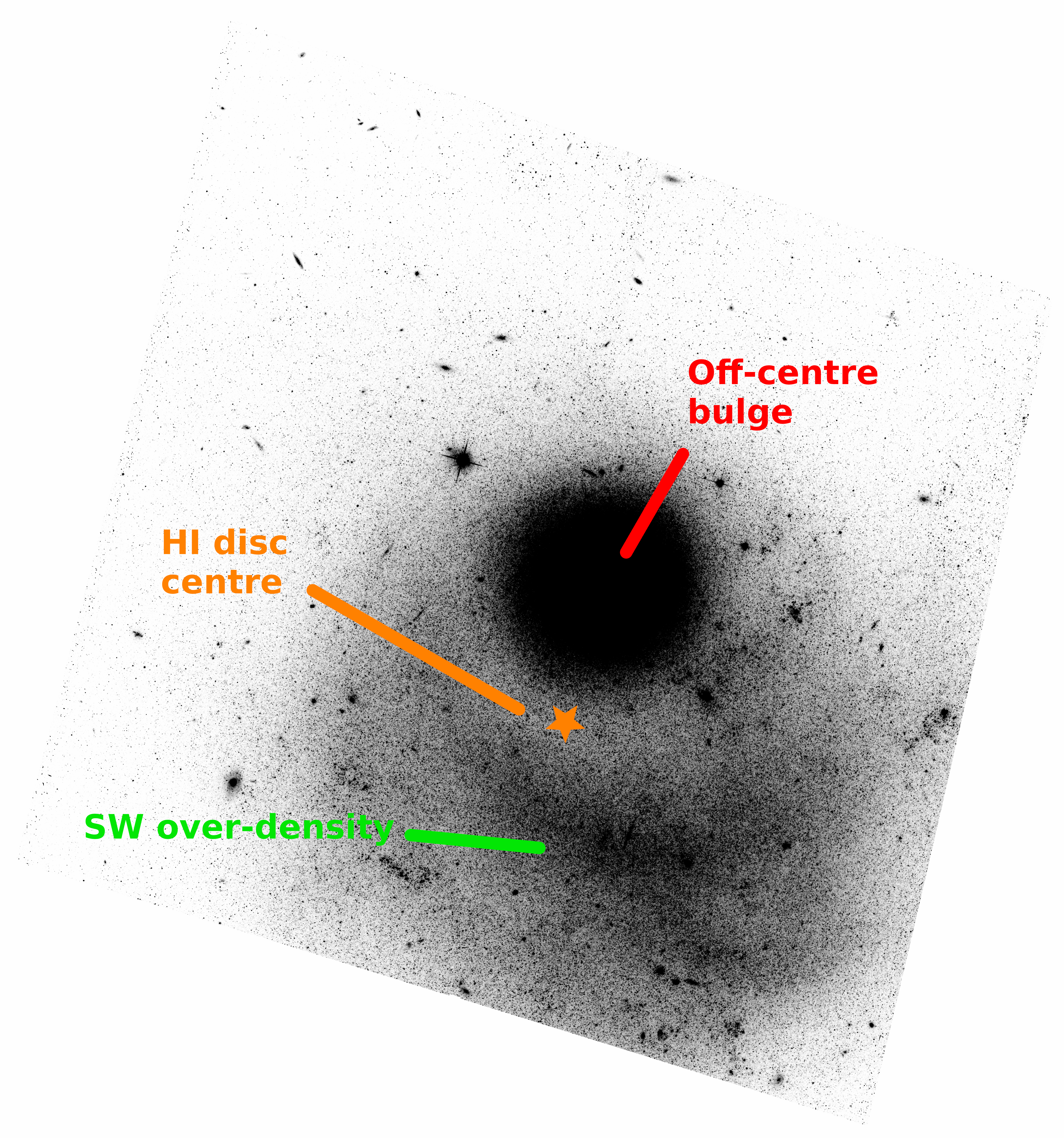}
	\caption{\textit{Left panel.} HST/ACS colour-combined image of NGC 5474 (blue = F606W, red = F814W). \textit{Right panel.} HST/ACS image of NGC 5474 in the F814W filter. The optical bulge of the galaxy is indicated in red, the South-West stellar over-density in green, while the orange star marks the kinematic center of the HI disc. In both panels North is up and East is to the left.\label{fig:ngc5474_image}}
\end{figure*}

In this paper we study the stellar populations and SFH of the dwarf galaxy NGC 5474, observed as part of the HST Treasury program LEGUS \citep{Calzetti2015}. NGC 5474 is a star forming dwarf spiral galaxy located at a distance of $\sim 7.0$ Mpc \citep{Tully2013,Sabbi2018} from the Milky Way and at a projected distance of $\sim 90$ kpc \citep{Tikhonov2015} from the giant spiral galaxy M101 (also known as the Pinwheel Galaxy). Its relatively nearby distance and the presence of M101 make this system an excellent study case  to explore the importance of environmental mechanisms in shaping the evolution and the SF activity in dwarf galaxies. The left panel of Figure \ref{fig:ngc5474_image} shows an HST/ACS colour-combined image of NGC 5474 (blue= F606W, red= F814W), while the right panel shows the F814W filter alone. The first thing that strikes the eye is a bright and compact bulge-like structure (marked in red in the right panel) located in the northern edge of the galaxy's face-on disc \citep{Kornreich2000}.\par

NGC 5474 was first observed in HI synthesis by \cite{Huchtmeier1979}, and subsequently, by \cite{Rownd1994} and \cite{Kornreich2000} with the Very Large Array (VLA). These observations show a smooth and fairly symmetric HI distribution that extends beyond twice the optical radius of the galaxy. The HI disc displays a very regular velocity field in the central region, while a pronounced warp develops at radial distances larger than $\sim 5$ kpc, and connects the gaseous component of NGC 5474 to the south-western edge of M101. Even more interesting, all previous investigators detected an off-set of $\sim 1$ kpc between the kinematic center of the HI disc (marked in orange in Figure \ref{fig:ngc5474_image}) and the position of the optical bulge. The right panel of Figure \ref{fig:ngc5474_image} reveals another striking feature of NGC 5474: the presence of a significant stellar over-density of the disc to the South-West of the bulge (hereafter \textquotedblleft SW over-density\textquotedblright, for brevity). The latter feature does not coincide with the spiral arms, but it seems to match the position of a local minimum in the HI distribution \citep{Bellazzini2020}. Historically, the many dynamical disturbances displayed by NGC 5474 have been attributed to its interaction with M101. The only attempt to model a possible close encounter between M101 and NGC 5474 by means of $N$-body simulations has been recently presented by \cite{Linden2022}. However, although the model that reproduces the properties of M101's disc involves a close passage of NGC 5474 through M101's outer disc $\sim 200$ Myr ago, the authors do not attempt to reproduce the highly asymmetric features observed in NGC 5474.\par 

The odd inconsistency between the HI kinematic center and the off-set position of the optical bulge of the galaxy has driven several hypotheses about the true nature of such a stellar component. \cite{Fisher2010} suggest that NGC 5474's bulge shows properties more similar to a pseudobulge than to a classic bulge (see \cite{Kormendy2004} for a comprehensive review). Indeed, this component exhibits some characteristics more similar to those of discs than classical elliptical bulges: recent star formation (see below), traces of spiral structure, and a surface brightness profile that is better described by an exponential law rather than by the classical $r^{1/4}$ de Vaucouleurs profile.  An interesting hypothesis that the bulge could be a satellite of NGC 5474, first put forward by \cite{Rownd1994}, has been recently revived in the literature by \cite{Mihos2013} and \cite{Bellazzini2020}. In particular, \cite{Bellazzini2020} suggest that its structural parameters are similar to those of a dE, in terms of stellar mass, V-band absolute magnitude, and half-mass radius. \cite{Pascale2021} explored this hypothesis by means of $N$-body hydrodynamical simulations, showing that a close encounter between an early-type compact dwarf galaxy and NGC 5474 might explain the peculiar features displayed by this galaxy (e.g., bulge-disc offset, SW-overdensity). The goal of this paper is to try to put further constraints on this hypothesis, recovering for the first time the SFH of NGC 5474.   

The paper is structured as follows. The LEGUS HST data used in this work are presented in Section \ref{data}. In Section \ref{stellar_populations} we present the colour-magnitude diagram and the spatial distribution of NGC 5474 main stellar populations. To quantitatively investigate the SFH we applied a new synthetic CMD routine, named SFERA 2.0, an evolution of the population synthesis routine SFERA (Star formation Evolutionary Recovery Algorithm), already used in previous works \citep[see e.g.,][]{Cignoni2015,Cignoni2016,Cignoni2018}. The method used to extract the SFH from the observed CMD is briefly outlined in Section \ref{SFH_recovery_method} (see Appx \ref{app:SFERA2.0} for more details).
We present the results in Section \ref{SFHs}, and conclusions are in Section \ref{Discussion and conclusions}.
 
\section{Data} \label{data}

In the following section we present the data we used in this work, together with the artificial star tests (ASTs) which we performed to estimate the photometric errors and incompleteness of the data.

\subsection{Photometry of resolved stars} \label{Photometry of resolved stars}
We retrieved from the LEGUS website\footnote{https://legus.stsci.edu/} the F606W (hereafter $V$) and the F814W (hereafter $I$) photometric catalogs for NGC 5474. These data were acquired with the Advanced Camera for Surveys Wide-Field Channel (ACS/WFC), as part of a multiwavelength photometric survey of 50 nearby galaxies of different morphological type within $\sim 12$ Mpc. Positions and fluxes for point-like sources were measured via PSF-fitting using the ACS modules of the photometry package DOLPHOT \citep{Dolphin2000,Dolphin2016}. All the details on the data reduction and on the parameters included in the catalogs are reported in \cite{Calzetti2015} and \cite{Sabbi2018}. The choice of the $V$ and $I$ filters is ideal to detect young massive stars as well as old ones (mostly represented by red giant branch stars, RGB), in order to be able to infer the SFH over the entire Hubble time.\par 
To clean the photometric catalog from spurious objects, we applied selections cuts on the DOLPHOT parameters \textit{MAGER} (magnitude error), $\chi^{2}$, \textit{SHARP} (object sharpness), \textit{CROWD} (object crowding), and \textit{ROUND} (object roundness) (see \cite{Dolphin2000,Dolphin2016,Sabbi2018} for a better description of these parameters). In particular, to retain a star, we required the following conditions: \textit{MAGER} $< 0.4$, \textit{CROWD} $< 0.2$, $\mid$\textit{SHARP} $\mid < 0.2$, $\chi^{2} < 2$, and \textit{ROUND} $< 3$ in both $V$ and $I$ filters. A visual inspection in the images of the rejected objects, coupled with an analysis of their radial profiles, revealed that the majority of them are foreground stars, unresolved stellar clusters, background galaxies, or blends of two or more stars. After these selections, we ended up with a catalog of about $232,000$ stars. 

\subsection{Incompleteness and photometric errors} \label{Incompleteness and photometric errors}
To estimate the photometric errors and incompleteness of the data, we performed ASTs on the images. ASTs mimic the observational process by injecting fake sources, generated with the appropriate Poisson noise and PSF, onto the images. To efficiently compute the completeness without altering the actual crowding of the stellar field, artificial stars are placed one at a time following the galaxy's light profile, covering the dataset's entire range of positions and magnitudes. At this point, the source detection routine (DOLPHOT) is applied to the field containing real and fake stars. An input artificial star is considered `lost' when it is not recovered in the output catalog, or when it is recovered more than 0.75 magnitude brighter. Indeed, in this latter case, the artificial star is likely blended with one or more other real stars, creating an `altered' source with more than twice the flux of the input one. We then further cleaned the artificial star catalog applying the same selection criteria on the DOLPHOT parameters used to clean the real data (see Section \ref{Photometry of resolved stars}). In total, we simulated $\sim 1.5$ million artificial stars. The ratio between injected and recovered stars in each colour and magnitude bin provides a measure of the completeness level, while the difference in magnitude between the input artificial stars and the recovered ones is used to estimate the photometric errors.\par

\begin{figure}
	\centering
	\includegraphics[width=\columnwidth]{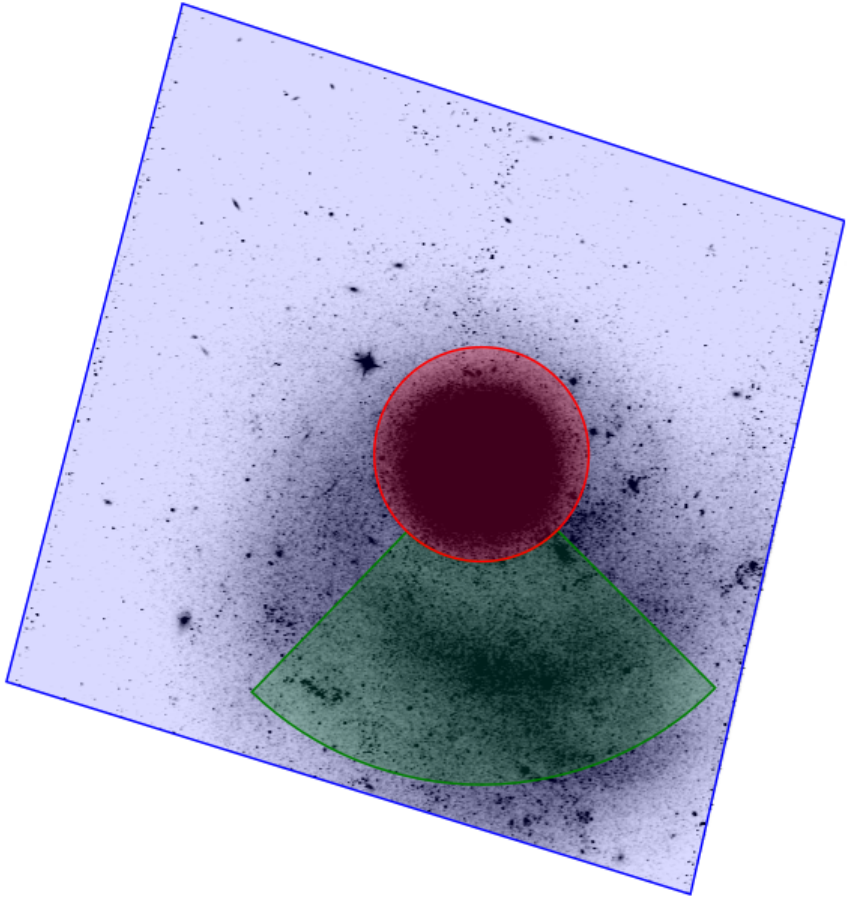}
	\includegraphics[width=\columnwidth]{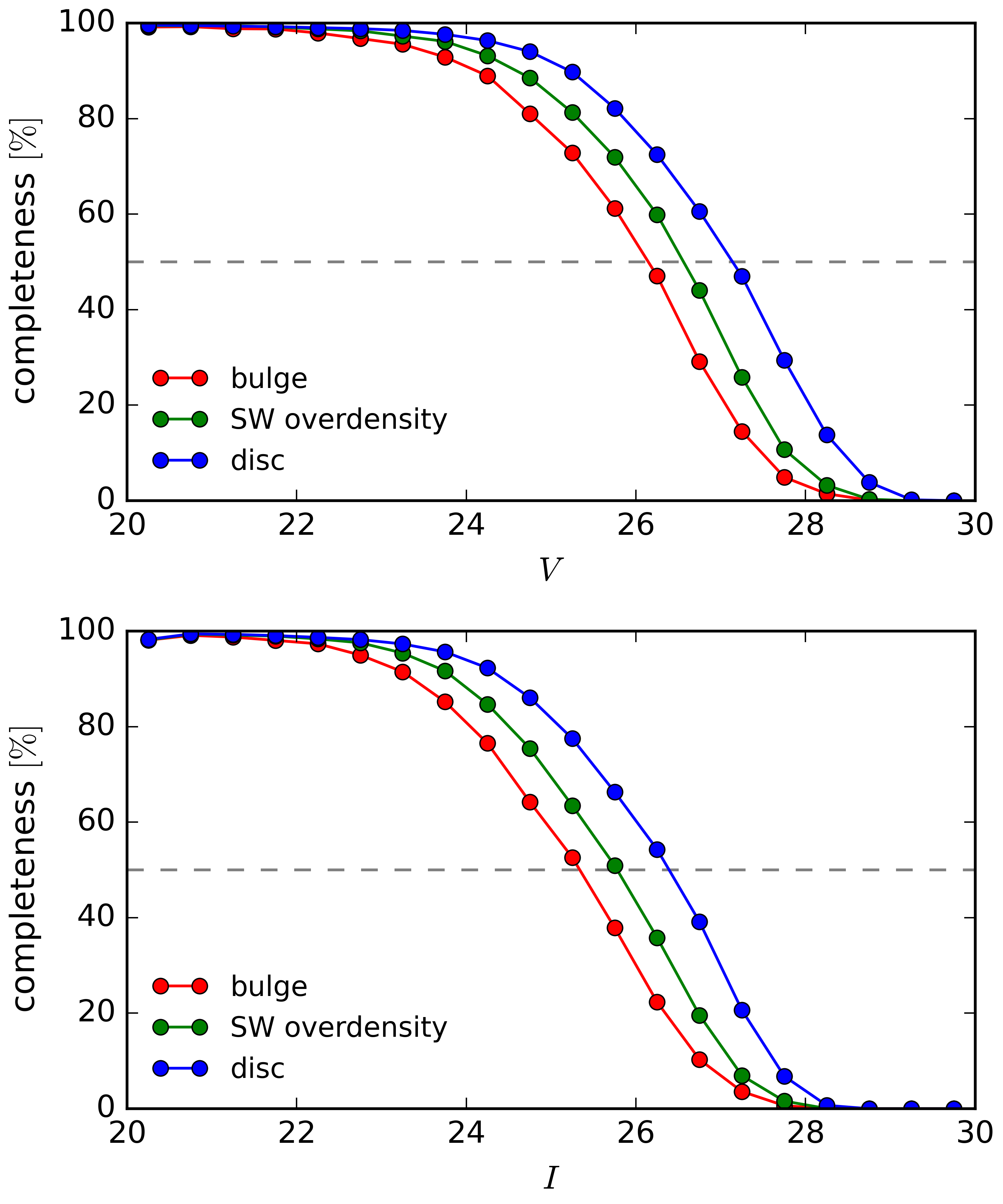} 
	\caption{\textit{Upper panel}: three regions selected in NGC 5474 to include: the bulge-like structure (red), the SW over-density (green), and the stellar disc (blue). \textit{Middle panel}: Average photometric completeness as a function of $V$ magnitude estimated from the ASTs in the three different regions outlined in the upper panel: bulge (red dots), SW over-density (green dots), and stellar disc (blue dots). The dashed line marks the 50\% completeness value. \textit{Lower panel}: the same as middle panel but for the $I$ filter. \label{fig:completeness_map}}
\end{figure}

Figure \ref{fig:completeness_map} shows the average completeness level in $V$ and $I$ for different magnitude bins, in three different regions of the galaxy, i.e. the optical bulge (red), the SW over-density (green), and the outer part of the stellar disc (blue). As expected, the bulge region is the most incomplete one due to extremely high crowding, with the completeness starting to drop below $50\%$ at $V \sim 26$ and $I \sim 25$. On the other hand, the outer part of the stellar disc, where stars are less packed against each other, shows a more than $50\%$ completeness limit down to $V \sim 28$ and $I \sim 27$.

\section{Stellar populations of NGC 5474}\label{stellar_populations}

In this section we analyse the colour-magnitude diagram of NGC 5474, and the spatial distribution of different stellar populations that trace different epochs.

\subsection{Colour-magnitude diagram}\label{colour-magnitude diagram}
\begin{figure}
	\centering
	\includegraphics[width=\columnwidth]{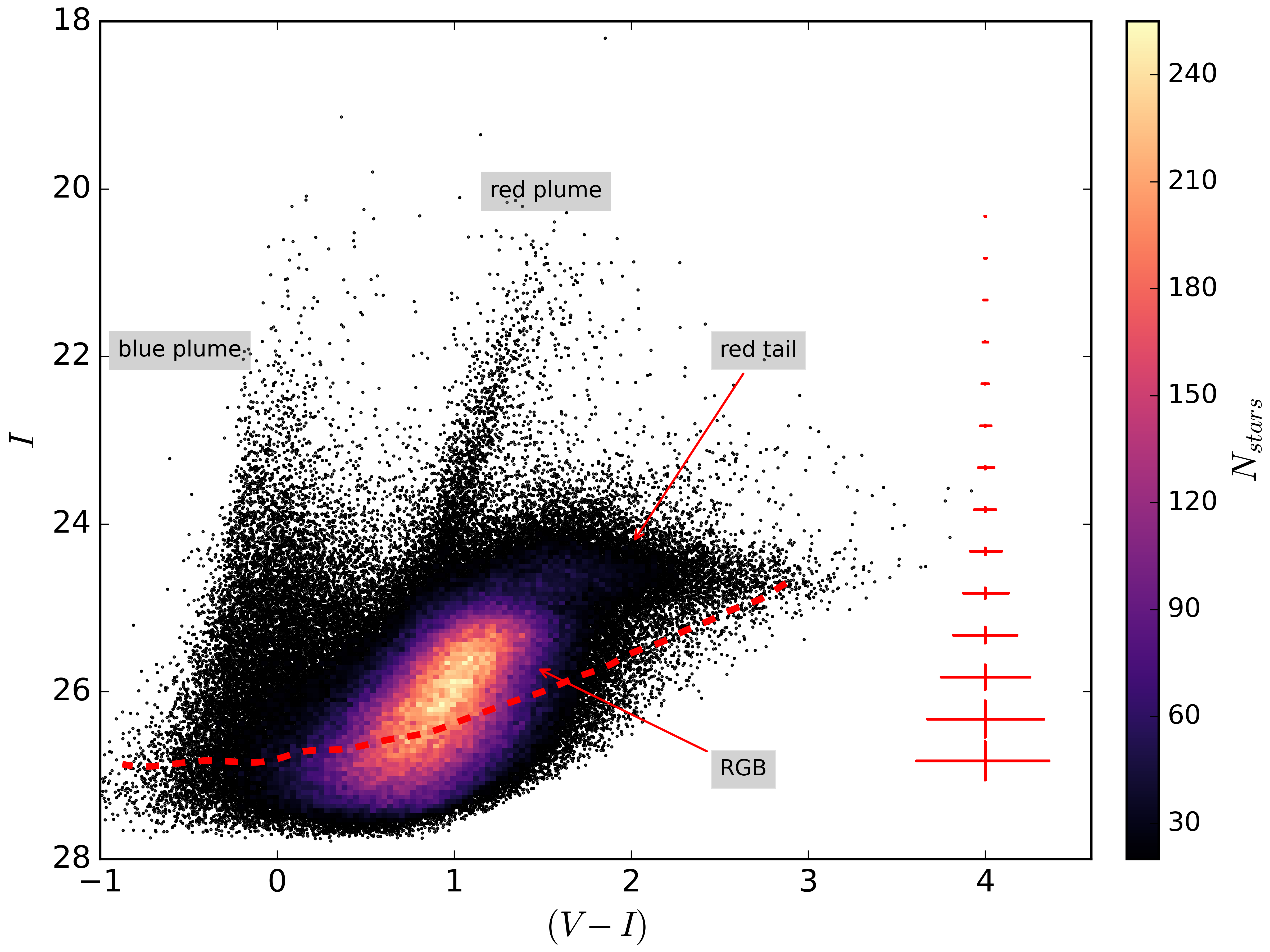} 
	\caption{CMD for the whole field of NGC 5474 covered by the HST ACS/WFC imaging (see Figure \ref{fig:ngc5474_image}), after applying all the photometry quality cuts described in Section \ref{Photometry of resolved stars}). The high density regions have been binned and colour coded in terms of density (see the colour bar on the right-hand side). The photometric errors shown in red on the right-hand side are those derived from the ASTs assuming $(V-I)=1$. The dashed curve represents the $50\%$ completeness level as derived from the ASTs. The main features discussed in Section \ref{colour-magnitude diagram} are labeled.\label{fig:CMD}
	}
\end{figure}

\begin{figure}
	\centering
	\includegraphics[width=\columnwidth]{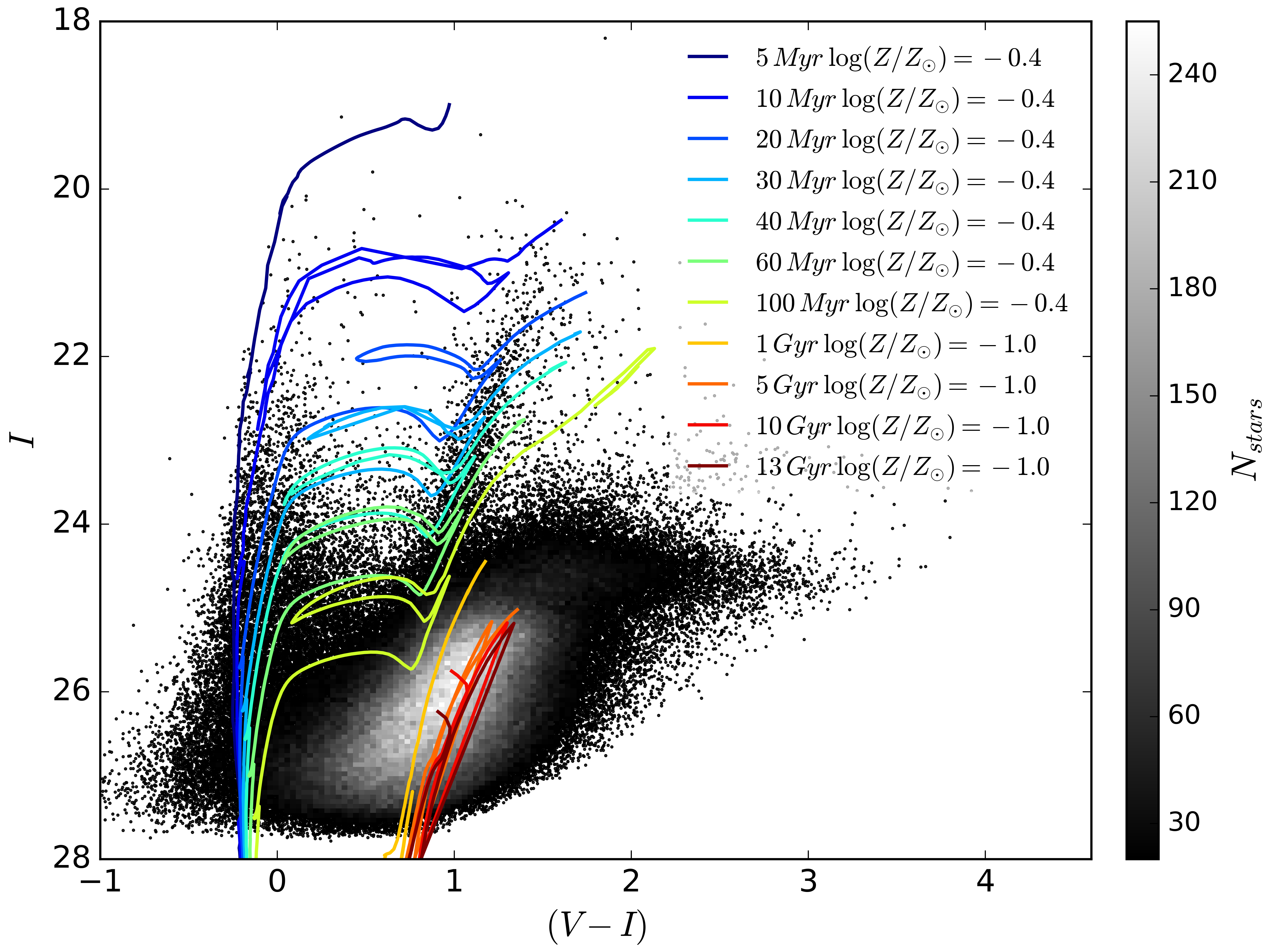} 
	\caption{CMD of NGC 5474 with superimposed isochrones with different ages and metallicities from the Padova PARSEC-COLIBRI database. The high density regions of the CMD have been binned and colour coded by the stars number density (see the colour bar on the right-hand side). We assumed a distance modulus of $(m-M)_{0}=29.22$ and a foreground reddening of $E(B-V)=0.01$. The isochrones are colour coded according to their age (see the labels on the top-right corner)\label{fig:CMD_plus_iso}. 
	}
\end{figure}

Figure \ref{fig:CMD} shows the $I$ vs. $(V-I)$ CMD of NGC 5474 covered by the HST/ACS WFC field, after applying the photometry quality cuts described in Section \ref{Photometry of resolved stars}. Photometric errors in different magnitude bins, as derived from the AST, are shown on the right-hand side. Also shown is an estimate of the $50\%$ completeness limit, represented by the red dashed curve. To aid the visualization, the high density parts of the CMD have been colour-coded according to the stellar density (see the colour bar on the right-hand side).
After a first qualitative inspection of NGC 5474's CMD, we can already identify a large variety of stellar populations, and recognize the main features of CMDs typical of star-forming dwarf galaxies (see labels in Figure \ref{fig:CMD}):
\begin{itemize}
	\item most of the stars are tightly clustered into a well-defined red giant branch, extending from $(V-I)\sim 0.5$ at $I\sim27$ to a clear tip located at $I\sim25$ and $(V-I)\sim 1$;
	\item above the RGB tip, a broad sequence of bright asymptotic giant branch (AGB) stars bending to the right and culminating in a `red tail' at $I\sim24.5$;
	\item a vertical and prominent blue plume, with average colour $(V-I)\sim 0$, hosting both upper-main sequence (MS) stars and stars at the blue edge of the core He-burning phase (the so-called blue loop, BL);
	\item a tilted (to the right) red plume that starts at $(V-I)\sim0.8$ and $I\sim24$ and extends up to $I\sim 20$, likely populated by a mix of He-burning stars near the red edge of the BL phase and early AGB stars of intermediate mass.  
\end{itemize}
To better identify the CMD regions corresponding to different stellar ages, we superimposed to the CMD a set of PARSEC-COLIBRI isochrones \cite[see e.g.,][]{Bressan2012,Chen2014,Tang2014,Chen2015,Marigo2017,Pastorelli2019} \footnote{Web interpolator available at 'http://stev.oapd.inaf.it/cmd'}, for the labelled ages and metallicities reported in Figure \ref{fig:CMD_plus_iso}. We assumed a distance modulus of $(m-M)_{0}=29.22$ \citep{Tully2013} and a foreground galactic extinction of $E(B-V)=0.01$ \citep{Schlafly2011}. For the youngest population ($\leq 100$ Myr, shown in Figure \ref{fig:CMD_plus_iso} in bluer colours), we used isochrones with metallicities compatible with the oxygen abundance measured in the H II regions by \cite{Moustakas2010}, namely $\log(Z/Z_{\odot})\sim -0.4$ (assuming $Z_{\odot}=0.0152$, \cite{Caffau2011}). The oldest isochrones ($\geq 1$ Gyr) have been chosen with $\log(Z/Z_{\odot}) = -1.0$, a metallicity for which the isochrones seem to well match the average RGB colour. The comparison with the isochrones reveals how the RGB is populated by stars older than $1-2$ Gyr, and possibly as old as the Hubble time. This means that these stars can be used to probe the average intermediate to ancient SF activity of the galaxy. On the other hand, all the stars lying in the blue and red plumes are younger than $100$ Myr and the signpost for the galaxy's recent SF activity. Finally, the AGB stars on the horizontal `red tail' trace both intermediate and old age populations ($\gtrsim 100$ Myr). Notice that, for the sake of clarity, in the 100 Myr and 1 Gyr isochrones the thermally pulsating AGB phase (TP-AGB) is not included in the figure. However, the TP-AGB phase is properly included in the CMD simulations (see Appx \ref{app:SFERA2.0}).  

\subsection{Spatial distributions}\label{spatial distribution}

\begin{table}
\centering
\begin{tabular}{c|c|c|c}
\hline
 \multicolumn{4}{|c|}{Relative star counts $[\%]$} \\
\hline
 Stellar population &Bulge& SW over-density & Disk\\
 \hline
 MS   & $0.6$    &$0.4$&  $0.4$\\
 BL   & $5.0$    &$1.9$&   $1.8$\\
 AGB   & $10.2$    &$7.0$&   $3.8$\\
 RGB   & $5.0$   &$4.8$&   $3.0$\\
 
\hline
\end{tabular}
\caption{Percentage of stars in the four stellar populations selected in Figure \ref{fig:CMD_population_selections} (i.e. MS, BL, AGB, RGB), relative to the total star count present in each of the galaxy spatial regions selected in Figure \ref{fig:completeness_map} (i.e. bulge, SW over-density, disk). In other words, if stars not included in any of the four selected CMD regions were also included as a fifth row, each of the three columns would total to $100\%$.}
\label{tab:table1}
\end{table}

\begin{figure}
	\centering
	\includegraphics[width=\columnwidth]{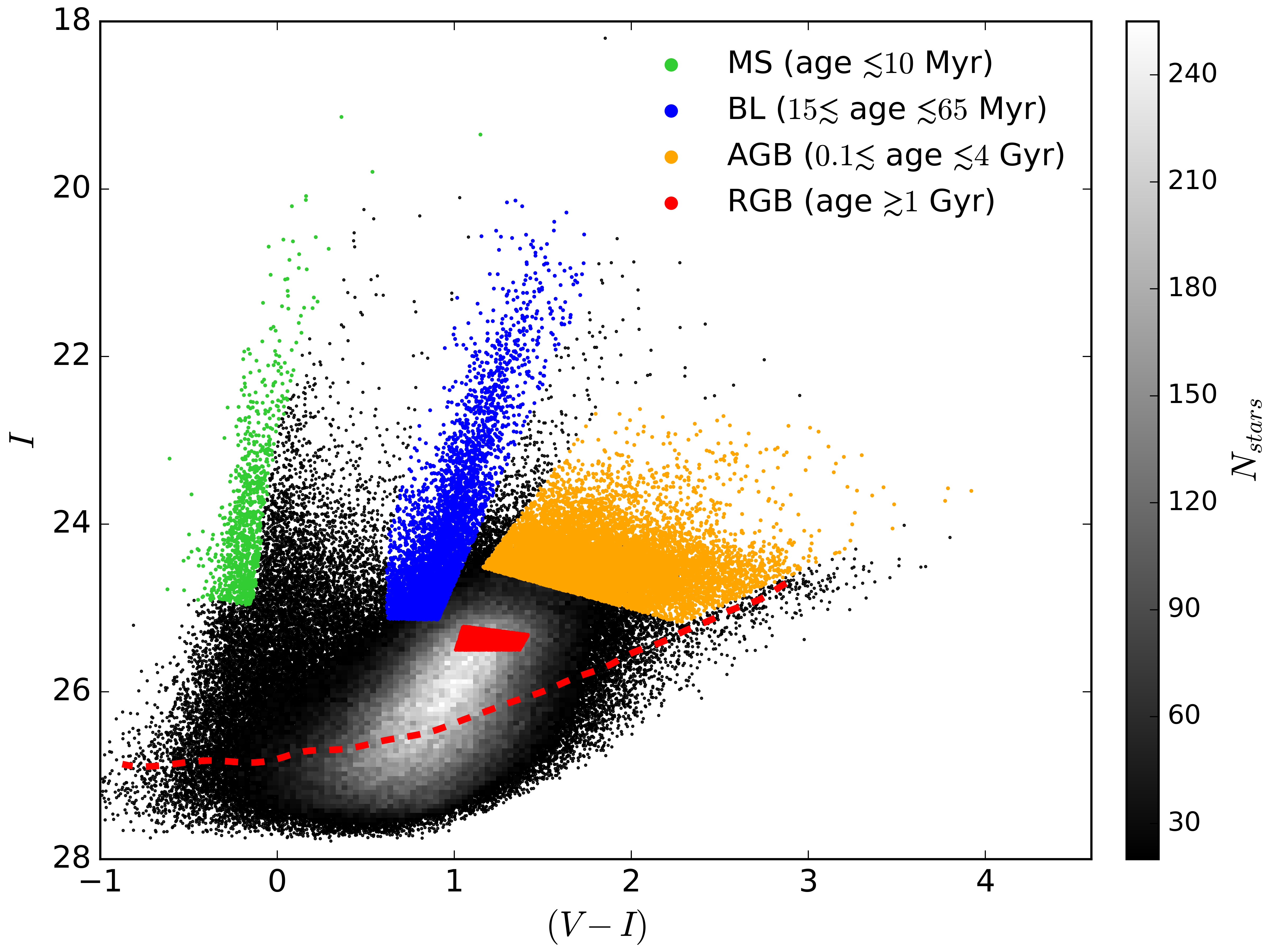}
	\caption{Selection in the CMD of four different stellar populations observed in NGC 5474, colour coded based on their age and evolutionary phase (see the legend in the upper right corner). In green, upper MS stars with ages $\lesssim 10$ Myr; in blue, He-burning stars at the red edge of the BL phase with ages between $\sim15$ and $\sim65$ Myr; in orange, AGB stars with ages between $\sim100$ Myr and $\sim4$ Gyr; in red, RGB stars tracing epochs older than $\sim1$ Gyr and as old as the Hubble time. The red-dashed curve shows the estimate of the $50\%$ completeness limit.}
	\label{fig:CMD_population_selections}
\end{figure}

In this Section we discuss how stars of different ages are distributed across the field of NGC 5474. For this goal, we selected stars in different evolutionary phases (see Figure \ref{fig:CMD_population_selections}), as tracers of different age intervals. Upper MS stars, tracing the last 10 Myr, are reported in green; He-burning stars at the red edge of the BL and early AGB stars of intermediate mass, likely tracing ages between $\sim 15$ and $\sim 65$ Myr, are in blue; AGB and RGB stars (near the tip), tracing the $\sim 0.1-4$ Gyr age range and epochs older than 1 Gyr are coloured in orange and red, respectively (see also the legend in the upper right corner). In order to avoid CMD regions affected by strong incompleteness, we selected only stars brighter than $I\sim25.5$ for our spatial analysis. Indeed, in the most crowded bulge region, this magnitude cutoff corresponds, for stars with a  $V-I$ color of $\sim1$, to a completeness level as high as $50\%$.

\begin{figure}
	\centering
	\includegraphics[width=\columnwidth]{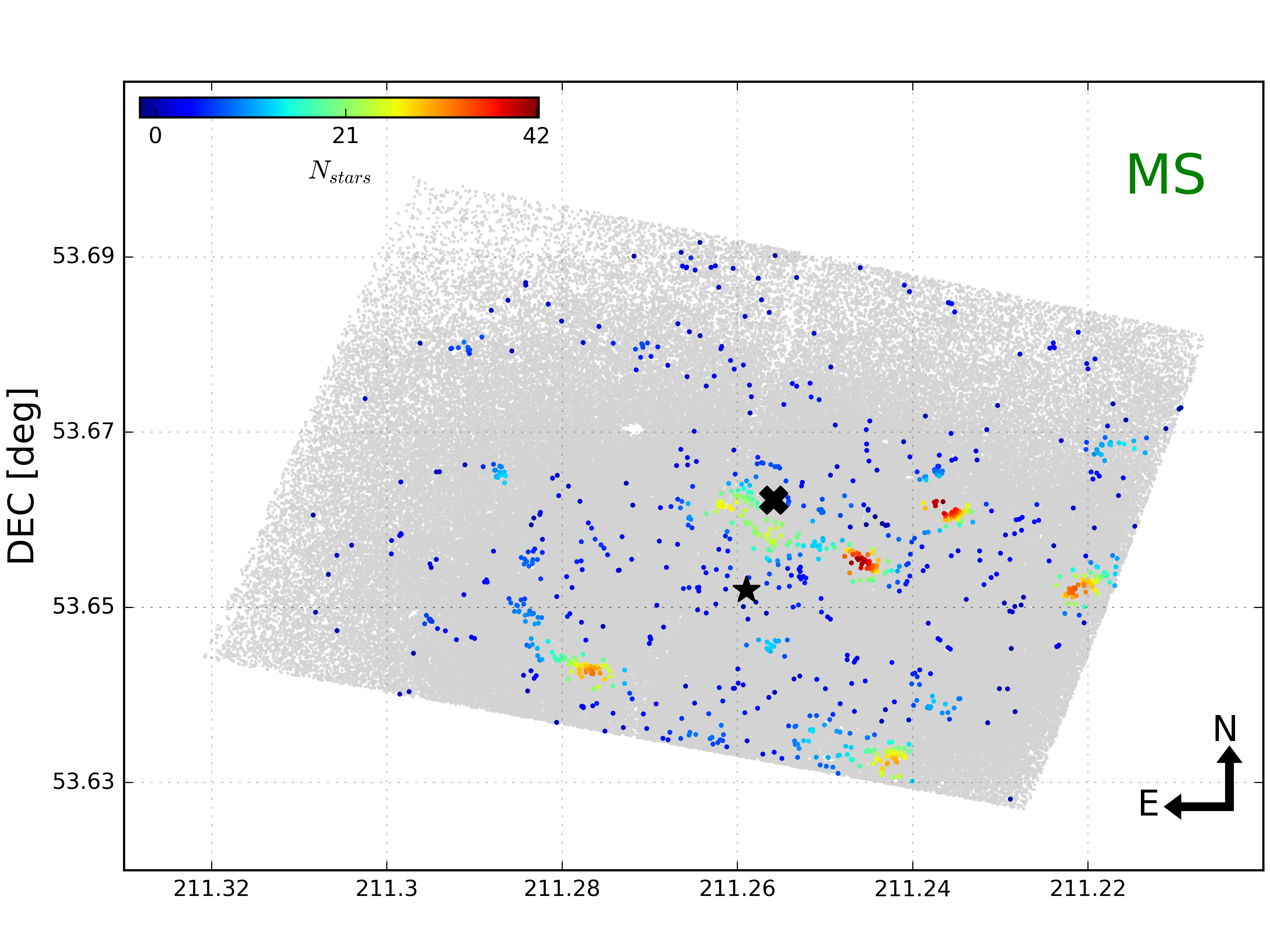}
	\includegraphics[width=\columnwidth]{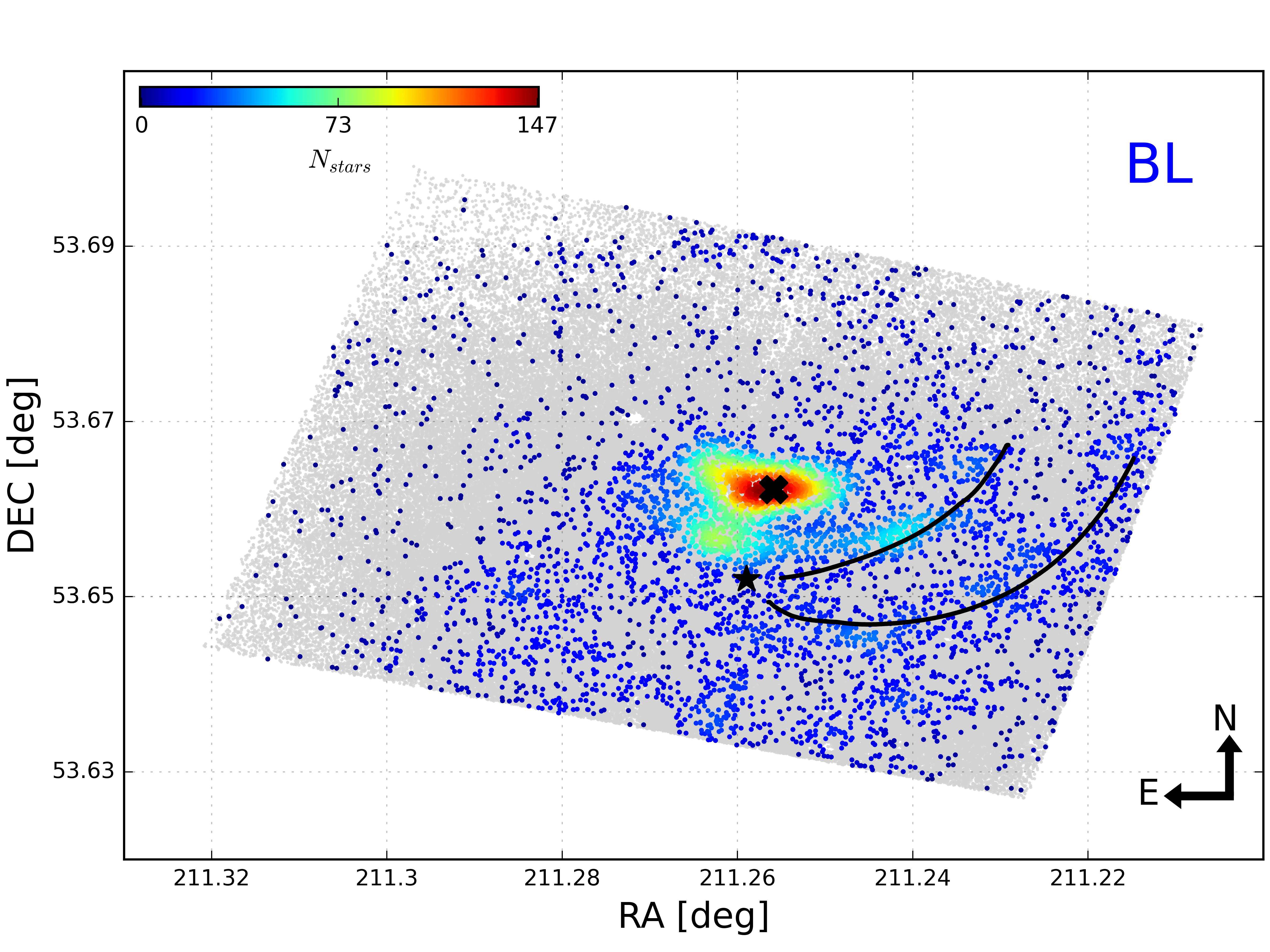} 
	\caption{\textit{Upper panel}. Density map of the MS stars shown in green in Figure \ref{fig:CMD_population_selections}. \textit{Lower panel}. Density map of the BL stars shown in blue in Figure \ref{fig:CMD_population_selections}. In both panels, colour changes according to the local density of stars (see the colourbar on the upper-left side of the panels), going from bluer colours (lower density) to redder colours (high density). The axis labels denote the position of the stars in RA and DEC. The black cross marks the position of the optical bulge, while the black star marks the kinematic center of the HI disc\label{fig:MS_BL_spatial_distr}. North is up and East is to the left. In the lower panel, the black-solid lines illustrate the 'tentative' position of the spiral arms.}
\end{figure}

Figure \ref{fig:MS_BL_spatial_distr} shows the spatial distributions of our bona-fide MS and BL stars, in the upper and lower panels respectively. The spatial distributions are binned and colour-coded according to the local density of stars (bluer colours indicate low density regions, redder colours high density ones), to better visualize their spatial features. The axis labels denote the position of the stars in RA and DEC. The black cross marks the position of the optical bulge, while the black star marks the kinematic center of the HI disc. \cite[see][]{Rownd1994}. In both panels, North is up and East is to the left. The spiral pattern is barely visible among MS stars (younger than $\sim10$ Myr), but some major concentrations with higher stellar densities are found in correspondence with the HII regions (see left panel of Figure \ref{fig:ngc5474_image}) and of the optical bulge. Although this might indicate that the bulge shows signs of recent star formation (see Section \ref{Bulge}), we cannot rule out a projection effect, due to part of the disc being in front of the bulge. On the other hand, BL stars (younger than $\sim100$ Myr) display a wide spiral pattern (illustrated by black-solid lines in the bottom panel of Figure \ref{fig:MS_BL_spatial_distr}), that extends through the entire field of view of the galaxy, eventually reaching the center of the bulge where stars are arranged in what seems to be a distorted oval shape, possibly a bar-like structure. Both the MS and the BL samples show an excess of counts in the South-West half of the galaxy.
\begin{figure}
	\centering
	\includegraphics[width=\columnwidth]{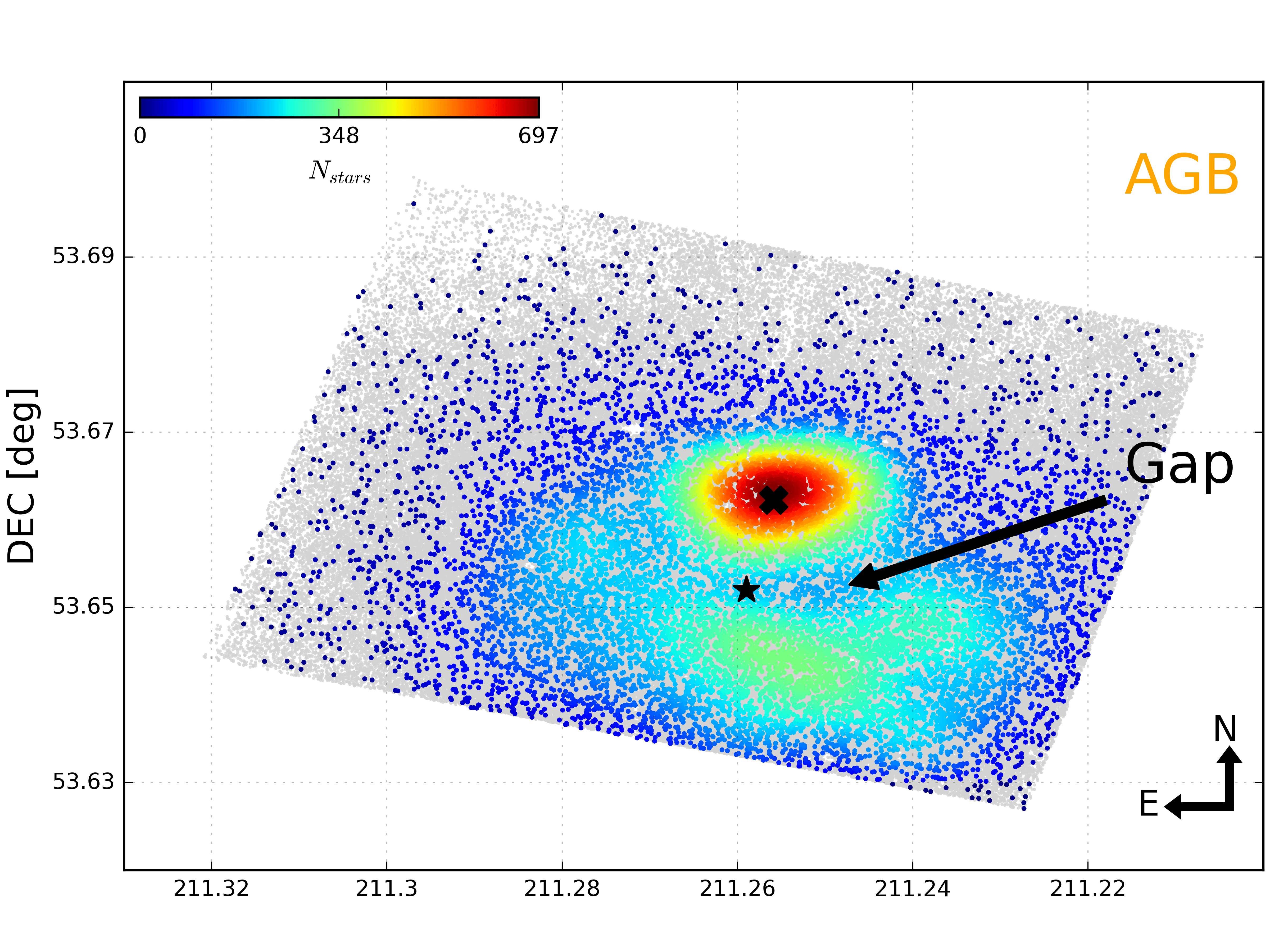}
	\includegraphics[width=\columnwidth]{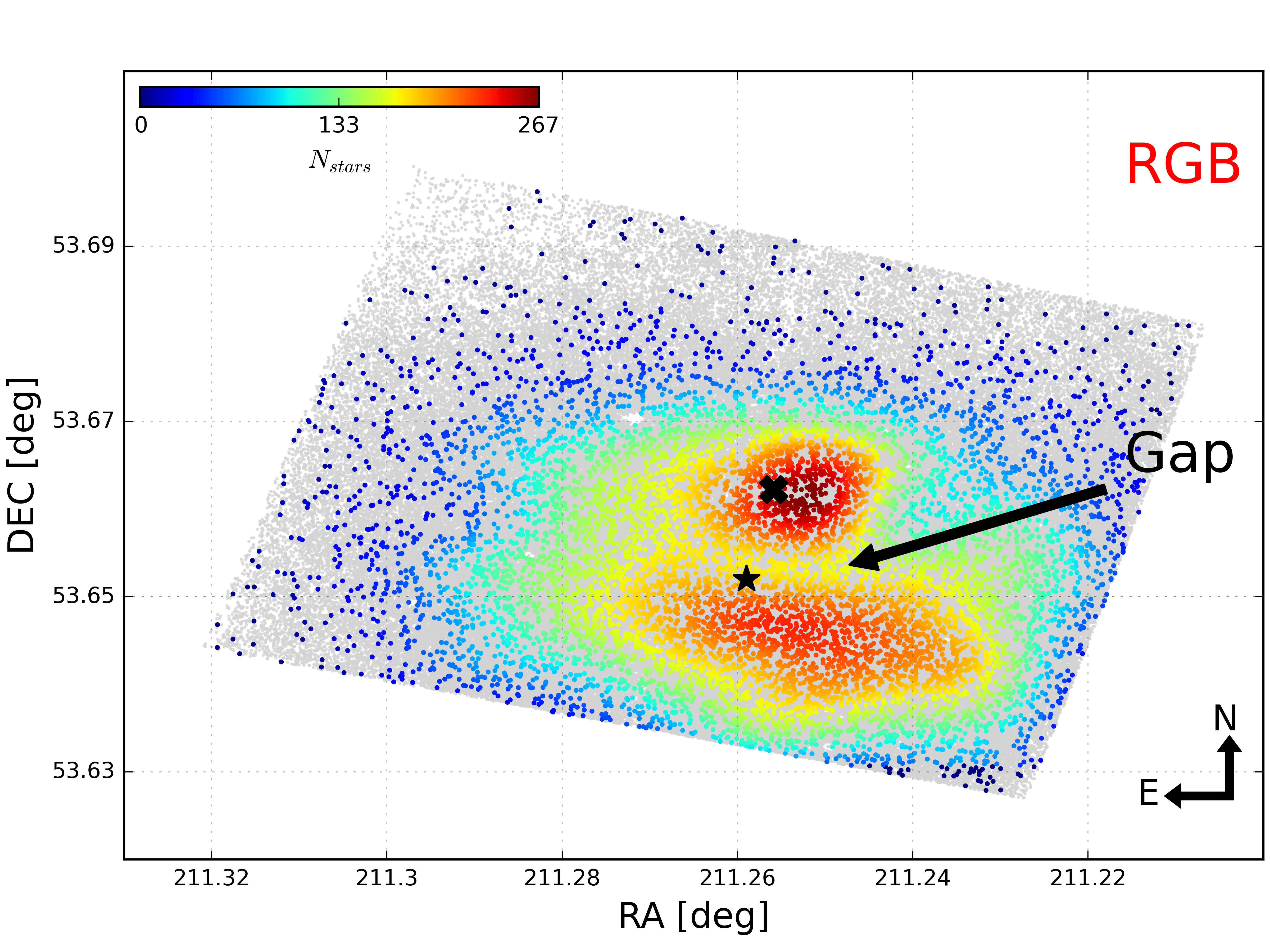} 
	\caption{\textit{Upper panel}. Density map of the AGB stars sample shown in orange in Figure \ref{fig:CMD_population_selections}. \textit{Lower panel}. Same for RGB stars. The black arrow  highlights a paucity of stars present in both AGB and RGB density map, producing a sort of bimodal distribution\label{fig:AGB_RGB_spatial_distr}. North is up and East is to the left.}
\end{figure}
The upper and lower panels of Figure \ref{fig:AGB_RGB_spatial_distr} show respectively the spatial distributions of the selected samples of AGB and RGB stars. Both the distributions are distinctly bimodal, with the densest component corresponding to the bulge and the secondary component to the SW over-density discovered by \cite{Bellazzini2020}. Clearly, both structures are dominated by intermediate and old age stars (see also Table \ref{tab:table1}). It is also worth noting that the paucity of stars  (hereafter the \textquotedblleft gap\textquotedblright , indicated in both panels of Figure \ref{fig:AGB_RGB_spatial_distr} by a black arrow) between the two structures almost coincides with the center of the HI distribution, and it is unlikely an artifact of the local incompleteness of the data. Indeed, both the AGB and RGB samples are above the 50\% completeness limit (with most of the regions above the 85\% completeness limit). Previous works \cite[see e.g.,][]{Bellazzini2020,Pascale2021} have already discussed the evidence of a gap between the bulge and the SW over-density. Here, thanks to our estimate of the incompleteness through extensive ASTs, we are able to corroborate this finding in a quantitative way. Of course, we cannot completely rule out the possibility that the gap is caused by a dust lane, as the ASTs would not have been able to trace it. However, the evidence that BL stars located in the gap do not show a significantly higher extinction seems to strengthen the hypothesis that the gap is real. \par

Furthermore, we measured the star counts for each of the stellar populations (i.e., MS, BL, AGB, RGB) and normalized them based on the total stellar count within each of the spatial regions selected in Figure \ref{fig:completeness_map} (i.e., bulge, SW over-density, disk). The outcomes of this analysis, presented in Table \ref{tab:table1}, substantiate the findings derived from the spatial maps discussed above. \par

The large asymmetries in the distribution of stars in NGC 5474 may have been produced by an inherently asymmetric star formation episode \cite[as already observed in other local dwarf galaxies, see e.g.,][]{Clementini2012,Rusakov2021}, or by a strong dynamical disturbance due to a gravitational interaction with one or multiple external galaxies \citep{Chonis2011,Fulmer2017}. As argued by \cite{Bellazzini2020}, the first hypothesis appears to be in conflict with the presence of a significant population of old stars ($\gtrsim 2$ Gyr) that comprise the SW over-density, as orbital mixing should have eliminated any over-densities caused by an asymmetric episode of star formation within the disc of NGC 5474 in just a few orbital times. Concerning the hypothesis of a dynamical disturbance, a possible candidate is the nearby giant spiral galaxy M 101. Indeed, according to $N$-body simulations, a close encounter between the two galaxies could have occurred $\sim 200$ Myr ago \citep{Linden2022}. This encounter may have created the asymmetries observed in the old stellar populations, rearranging the old AGB and RGB stars, and also triggered the more recent SF activity marked by the HII regions present along the spiral arm of the galaxy. As an alternative, the SW over-density could be the remnant of a disrupted satellite unrelated to M 101, whose merging brought old stars to NGC 5474 and possibly triggered the observed enhancement in the recent SF in the Southern half of the disc \citep[see e.g.,][]{Bellazzini2020, Pascale2021}.\par

Finally, the bulge itself could be a dwarf galaxy companion of NGC 5474, an hypothesis already discussed in the literature \citep[see e.g.,][]{Mihos2013,Bellazzini2020,Pascale2021}. In particular, by means of hydrodynamical $N$-body simulations, \cite{Pascale2021} explored two different scenarios for the possible nature of this component:
\begin{enumerate}
	\item a stellar system without dark matter, orbiting within the plane of the disc of the galaxy;
	\item an external compact early-type dwarf galaxy moving on a polar orbit around NGC 5474.
\end{enumerate}
In the former case, the authors found that an interaction of NGC 5474 with a putative bulge more massive than $\sim 10^{8}$ \MSUN would be hardly compatible with the smooth HI distribution and with the regular velocity field displayed by NGC 5474 \citep{Rownd1994,Kornreich2000}. Also, such a system would lose the majority of its mass, losing its spherical symmetry and developing visible tidal tails inconsistent with observations. Furthermore, this scenario does not explain the mechanism that may have caused the observed offset between the bulge and the HI kinematical center. In the second scenario of the putative bulge as a compact early-type dwarf galaxy satellite, the authors found not only that the off-set between its position and the center of the NGC 5474's disc can be easily produced by simple projection effects \citep[see Figure 10 in][]{Pascale2021}, but also that the gravitational interaction between the two systems could explain the warped HI distribution of NGC 5474 \citep{Rownd1994} and the formation of its loose spiral arms  \citep[see Figure 12 in][]{Pascale2021}.

In the next section we will present the method used to quantitatively recover the star formation histories of the main stellar populations (i.e., bulge, SW over-density, and disc) which form NGC 5474, in order to help put all the pieces of this galactic puzzle together. 

\section{SFH recovery Method}\label{SFH_recovery_method}
For this work, we developed a new Python-based SFH recovery routine (named SFERA 2.0). This routine is an improved version of the previous SFH recovery code SFERA \citep[see][for an extensive description]{Cignoni2009,Cignoni2015,Cignoni2016,Sacchi2018}. In this section we provide only a short description of SFERA 2.0 approach, while a more detailed analysis of the method can be found in Appendix \ref{app:SFERA2.0}, together with some benchmark's of the new routine.\par
The study of the SFHs of galaxies through CMD analyses involves comparing the observed stellar population CMD with a set of synthetic population CMDs created from stellar evolutionary models. Constraining the full functional forms of the SFH and the age-metallicity relationship (AMR) that best reproduce the observed CMD is a daunting task. A possible solution to this problem is to break down the complex history of a galaxy into `quasi-simple' stellar populations (QSSPs). Each QSSP is computed considering a population of synthetic stars with a uniform distribution of ages and metallicities within the intervals $\Delta t$ and $\Delta Z$, centered around $N$ discrete values of ages, and $M$ metallicities. Thus, within a QSSP the star formation rate (SFR) and metallicity are constant over time. Generating isochrones with these ages and metallicities, and populating them according to an initial mass function (IMF) and the stellar evolution times, transforms these QSSPs into synthetic CMDs. The CMD of any complex SFH/AMR can be approximated by a linear combination of these basis CMDs. In this way, the problem of inferring the most likely SFH/AMR behind the data translates into finding the combination of $N \times M$ basis CMDs that best reproduces the observational CMD. In order to find the SFH and AMR that most likely correspond to the observed galaxy's CMD, we need a reliable way to generate these galaxies' building blocks.\par
As a first step to generate all basic synthetic CMDs, we assumed the following input parameters:
\begin{itemize}
	\item a set of isochrones from the latest PARSEC-COLIBRI stellar models, which include all stellar evolutionary phases from the pre-main sequence to the thermally pulsing AGB phase \citep{Bressan2012,Marigo2017}; 
	\item a \cite{Kroupa2001} IMF between 0.1 and 150 \MSUN;
	\item a 30\% unresolved binary fraction \cite[see][for an extensive discussion of this particular choice]{Cignoni2009}
\end{itemize}
The synthetic basis CMDs are computed for ages from 1 Myr to 14 Gyr. From 1 Myr to 1 Gyr we use 12 equal logarithmic age bins of $\Delta\log(age) = 0.25$. The logarithmic step is chosen to take into account the decrease in the time resolution as the look-back time increases. This age range is traced in the CMD by bright stellar evolutionary phases (e.g., upper-MS, BL), which are good age indicators \citep{Gallart2005,Cignoni2010}, and easily detected at the distance of NGC 5474. Unfortunately, this cannot be said for epochs older than 1 Gyr. The only SF tracer of these epochs present in NGC 5474's CMD is the RGB, which is a poor age indicator \citep{Gallart2005,Cignoni2010}. For this reason, we opted for a single age bin from 1 to 14 Gyr. We do not assume any particular AMR. However, since we have information on the current metallicity of the galaxy from the nebular oxygen abundance derived from HII regions (hosting massive stars younger than $\sim 10$ Myr) corresponding to $\log(Z/Z_{\odot}) = -0.4 \pm 0.22$ \citep[see][]{Moustakas2010}, we allowed the metallicity of the models to vary in the range $\log(Z/Z_{\odot}) = [-2.0, -0.2]$ in 18 metallicity bins of $\Delta Z = 0.1$. Each synthetic basic CMD is generated with a uniform distribution in $Z$ within the metallicity bin $\Delta Z$. Thus, we ended up with a library of $13 \times 18$ synthetic basis CMDs.\par
The observational conditions were simulated convolving the `clean' synthetic basis CMDs with the following data properties: 
\begin{itemize}
	\item a foreground reddening of $E(B-V) = 0.01$ \citep{Schlafly2011}, and an extinction law with a normal total-to-selective extinction value of $R_{V} = 3.1$ \citep{Cardelli1989};
	\item a distance modulus of $(m-M)_{0} = 29.22 \pm 0.20$, obtained with the RGB tip method by \cite{Tully2013};
	\item photometric errors, incompleteness, and blending estimated from extensive artificial star tests (see Section \ref{Incompleteness and photometric errors}).
\end{itemize}
Finally, to compare models with data, `dirty' basis CMDs (i.e. convolved with the distance, reddening, photometric errors, and incompleteness of the data), as well as the observed CMD, are divided into a grid of magnitude and colour cells (producing Hess diagrams). In order to decide if a given linear combination of synthetic basis CMDs is a good representation of the data, the observed and the model star-counts in each grid cells must be compared. The minimization is implemented following Poissonian statistics, as already suggested by \cite{Dolphin2002}, exploring the entire age-metallicity space with the Pikaia genetic algorithm \citep{Charbonneau1995}\footnote{Routine developed at the High Altitude Observatory and available at the public domain: http://www.hao.ucar.edu/modeling/pikaia.php.}. The choice of a genetic algorithm, instead of a classical local hill climbing method, is dictated by the high number of free parameters (typically of the order of $10^{2}$), and the complex nature of the parameter space to be explored. For these reasons, a global optimization method must be used to explore the parameter space in more points simultaneously, and thus be independent of the initial conditions. Finally, to estimate the statistical uncertainties around the `best' model, we performed 30 bootstrap tests for each SFH analysis.

\section{Star Formation Histories}\label{SFHs}
\begin{figure}
	\centering
	\includegraphics[width=\columnwidth]{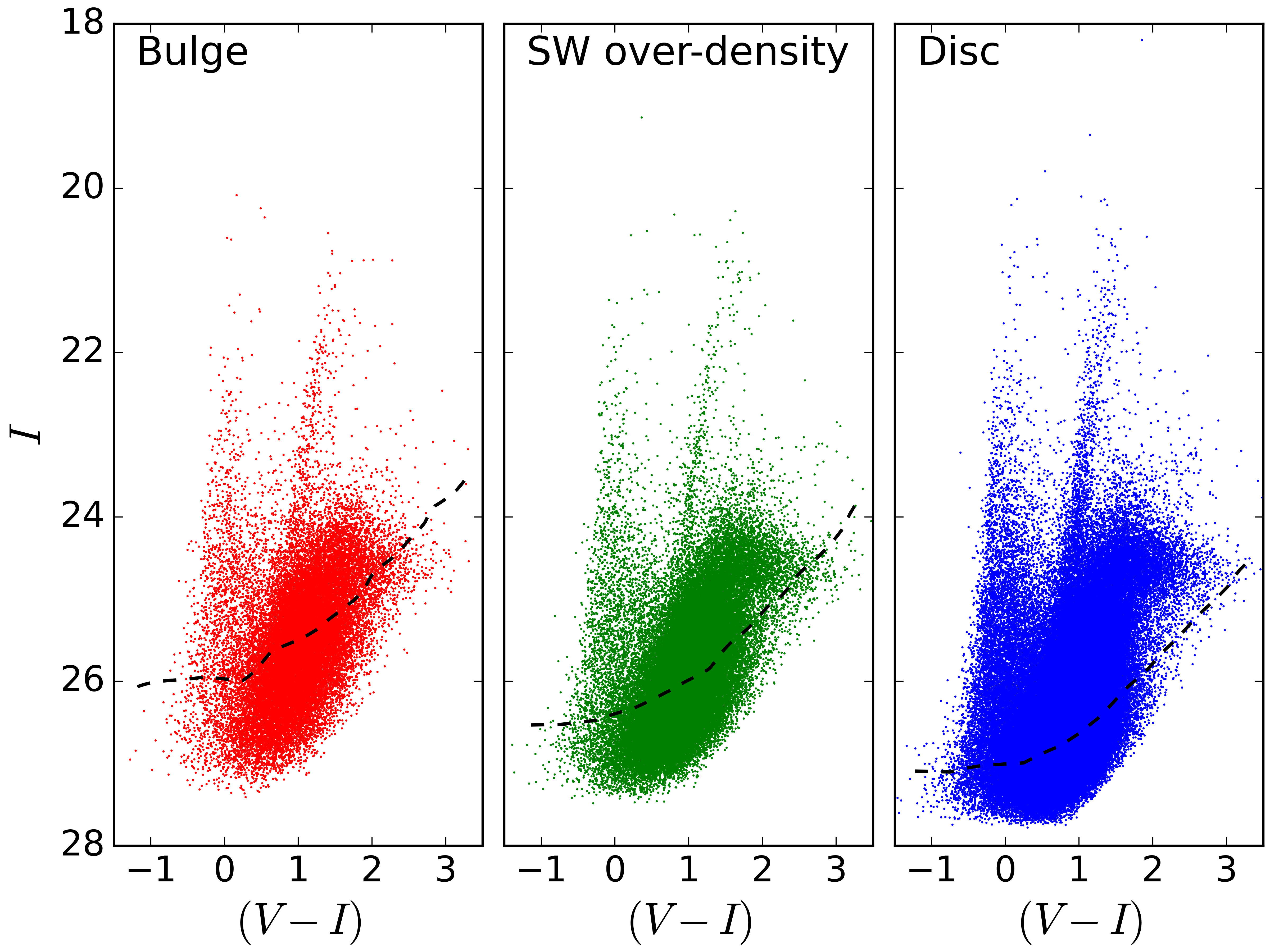} 
	\caption{The CMDs of the three regions into which we divided the whole NGC 5474 field (shown in top panel of Figure \ref{fig:completeness_map}). The optical bulge is in red, the SW over-density in green, and the disc in blue. The black dashed lines indicate the 50\% completeness limit estimated from artificial star tests. \label{fig:three_CMDs}}
\end{figure}

Given the rich and complex morphology of NGC 5474, we divided the whole galaxy field into three separate regions, for which we performed the SFH analysis independently. Figure \ref{fig:three_CMDs} shows the CMDs of the selected regions: the optical bulge in red, the SW over-density in green, and the outer field mainly composed of the galaxy's disc in blue (see the top panel in Figure \ref{fig:completeness_map} for the corresponding spatial maps). The CMD depth of the three selected fields clearly reflects the different crowding conditions of the sub-regions, with the bulge being the most crowded one. Indeed, the 50\% completeness level (black dashed line), estimated from ASTs, is reached at deeper magnitudes moving from the bulge to the outer part of the disc, as already discussed in Section \ref{Incompleteness and photometric errors}. Remarkably, all three CMDs show the presence of bright young stars (blue and red plumes at $(V-I) \sim 0$ and $(V-I) \sim 1$, and $I < 24$), together with prominent AGB and RGB phases, likely populated by intermediate and old age stars respectively. The simultaneous presence of multiple stellar evolutionary phases, both bright and faint, along with the relatively shallow photometric depth of the data (due to the galaxy's distance), makes the SFH recovery process rather challenging. Even more uncertain is the derivation of the galaxy's stellar metallicity. The upper-MS phase sampled by our CMD is fairly insensitive to metallicity variations. In contrast, the RGB, which could in principle provide hints about the overall metal content of the population older than $\sim2$ Gyr, exhibits a wide color spread, mainly due to high photometric errors at magnitudes fainter than $I \sim 25$.

\subsection{Bulge}\label{Bulge}
\begin{figure*}
	\centering
	\includegraphics[width=\textwidth]{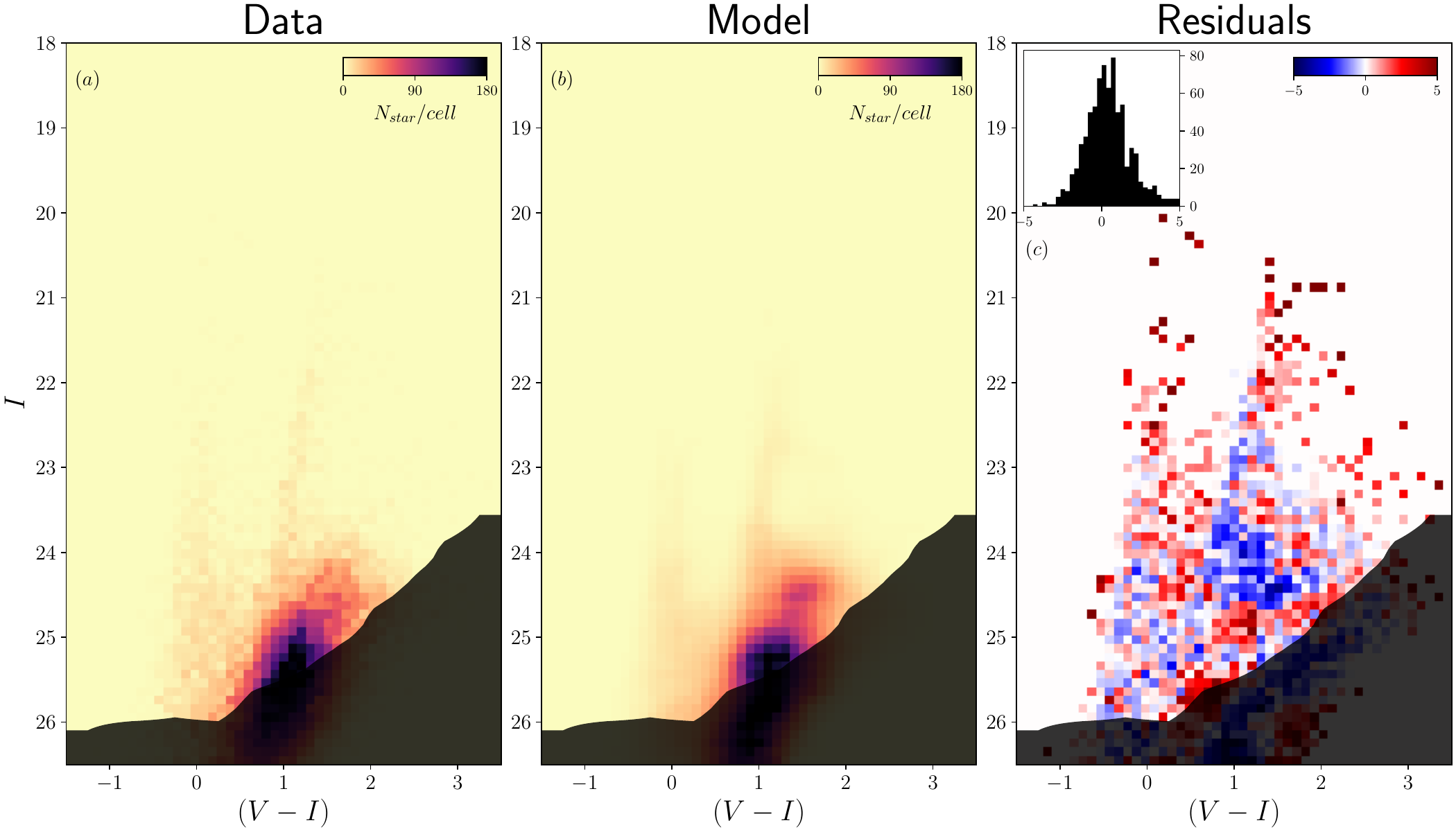}
	\includegraphics[width=0.49\textwidth]{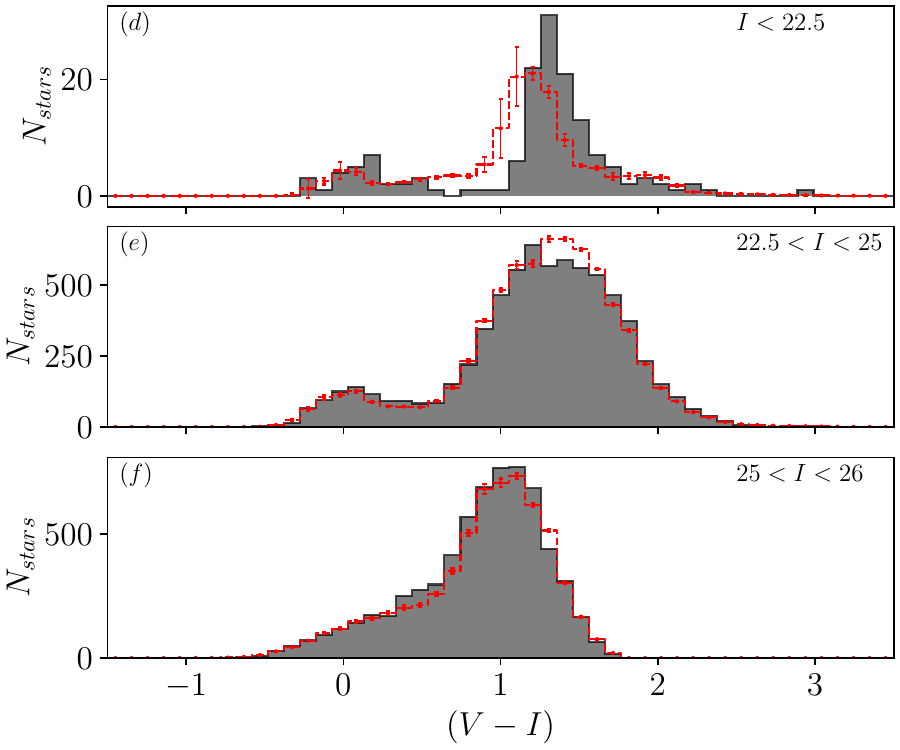}
	\includegraphics[width=0.49\textwidth]{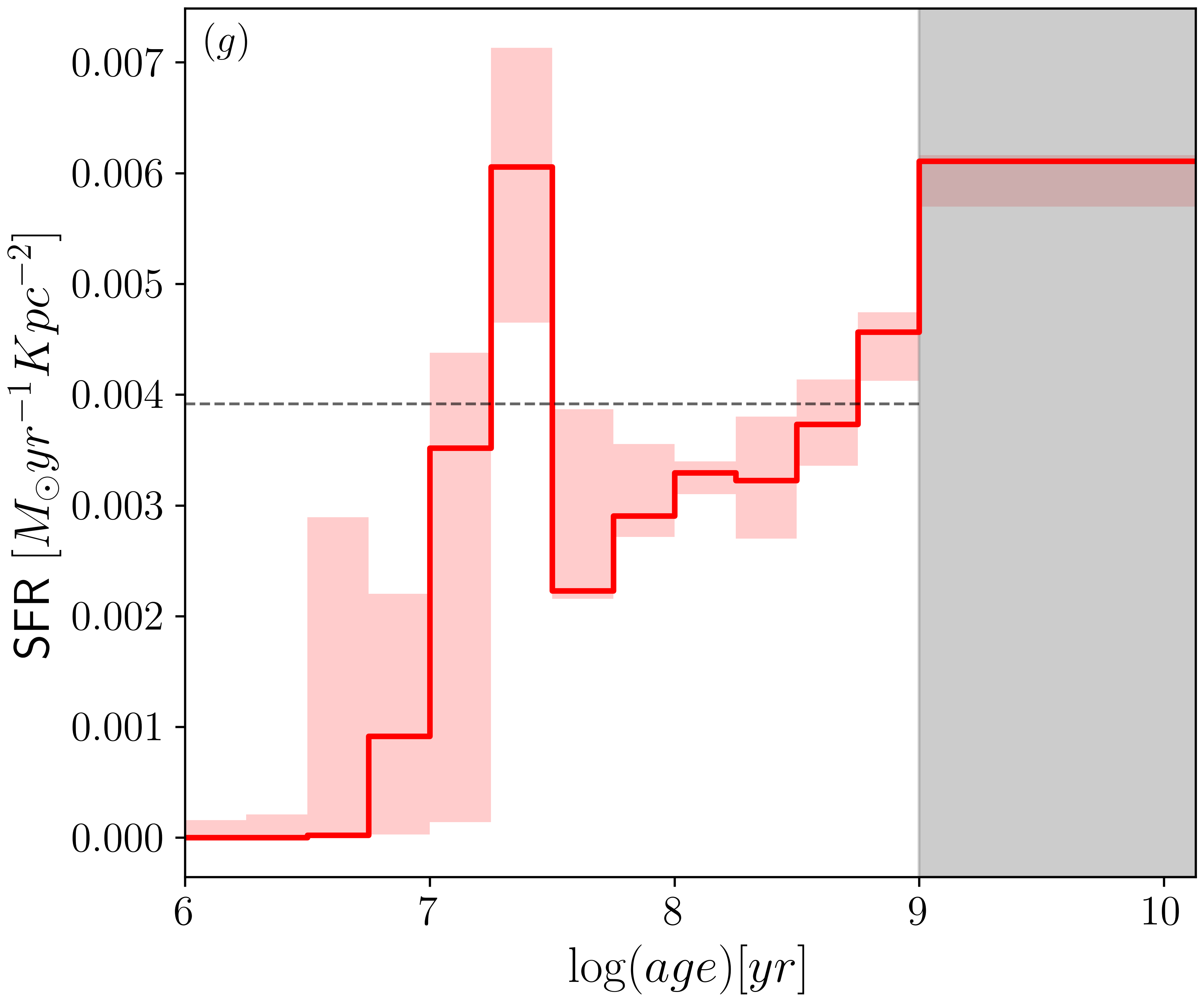}
	\caption{\textit{Top panels}. (a) Observational Hess diagram of the bulge; (b) best-fit synthetic Hess diagram reconstructed by SFERA 2.0; (c) the normalised residuals, with their distribution shown in the inset panel on the upper left. The shaded part of the diagrams corresponds to the region in the CMD below the $50\%$ completeness limit. \textit{Bottom panels}. (d)-(f) colour functions of the CMDs in different luminosity bins; the data are the gray histograms, while the model is indicated by the red dots with the uncertainties from the bootstrap routine. (g) Recovered SFH, normalised over the area of the region, using SFERA 2.0 and the PARSEC-COLIBRI stellar models. The gray-dashed horizontal line marks the mean SFR over the last Gyr. The epochs older than 1 Gyr are shaded due to the lack of reliable age indicators in the CMD.\label{fig:bulge_bestmodel}}
\end{figure*}

Figure \ref{fig:bulge_bestmodel} shows the results obtained applying our SFH recovery routine to the bulge of NGC 5474. The recovered SFH (panel (g)), normalised to the area of the region, shows the following features:
\begin{itemize}
    \item an overall decreasing trend in the last 1 Gyr;
    \item a very recent peak around 20-35 Myr ago (although characterized by rather large uncertainties, namely $\sim 30 \%$);
    \item an average SFR in the last 1 Gyr $\sim 1.5$ times lower than at older epochs. 
\end{itemize}

Overall, this result confirms that the bulge is a predominantly old population. Nevertheless, its SFH shows traces of recent SF activity, with no substantial gaps for the entire Hubble time, suggesting that the region has continuously been active until very recent times ($\sim 10$ Myr ago). As previously mentioned, this feature along with the 1 kpc off-set with the kinematic center of the HI distribution could corroborate the hypothesis that this structure is not a real bulge, but rather something else (e.g., a pseudobulge or an independent dwarf galaxy orbiting around NGC 5474).\par
Panel (a) of Figure \ref{fig:bulge_bestmodel} shows the observed Hess diagram of NGC 5474's bulge, together with the best-fit model (panel (b)), and the residuals in unit of Poisson uncertainty: $(data-model)/\sqrt{model}$ (panel (c)). We excluded from the minimization process the cells of the CMD where the completeness drops below $50\%$ (shaded area in the Hess diagrams), where the observational uncertainties become too large. As shown by the observational Hess diagram, our photometry reaches the tip of the RGB population but is not deep enough to identify  older features such as the red clump or the horizontal branch. As discussed in Section 4, this inevitably prevents us from precise age-dating beyond $\sim 2$ Gyr. Nevertheless, we are confident that this limitation does not affect the main conclusions of the analysis. \par

In principle, SFERA 2.0 allows us to infer also the AMR of the studied population. However, the inferred relation is robust only when the oldest MS turn-off is reached \citep{Gallart2005,Cignoni2010}. In our case, where the oldest measured stars are RGB, strongly affected by the well-known age-metallicity degeneracy \footnote{i.e., younger, metal rich RGB stars can have the same mag and colour than an older, metal-poor one.} the derived relation would be ill determined, and we prefer to limit our discussion to robust results.\par 

The code is able to reproduce well the overall morphology of the observed Hess diagram, although some differences are present. In particular, the residuals seem to suggest that the best model under-predicts the number of stars in the red plume at $(V-I)\sim 1.2$ and $I\lesssim 22.5$ (mainly populated by core He-burning stars of intermediate mass). Nevertheless, $90\%$ of the residuals are within $\pm3.0 \sigma$.\par
Panels (d)--(f) in Figure \ref{fig:bulge_bestmodel} show also the colour distribution of the data (gray histogram) and best-fit model (red dots) in different luminosity intervals ($I<22.5 $ panel (d), $22.5<I<25$ panel (e), and $25<I<26$ panel (f)). Although this kind of comparison is less informative than the residuals map, it allows us to focus on possible large discrepancies.  Again, panel (d) suggests that SFERA 2.0 under predicts the number of stars in the red plume ($0.5 \lesssim (V-I) \lesssim 1.5$). Moreover, it predicts a slightly bluer red plume than the observed one. Panel (e) also shows that our best model slightly overestimate the number of stars in the CMD region $22.5<I<25$ and $1 \lesssim (V-I) \lesssim 2$, populated by AGB and TP-AGB stars. Nonetheless, colour distributions are globally well-fitted by the model within the uncertainties.\par

To check the impact of the assumed distance on the SFH recovery, we re-derived the SFH by letting the distance free to vary in steps of 0.01 mag in the range $(m-M)_{0} = [29.02,29.42]$ given by the uncertainties affecting the distance measure \citep{Tully2013}. We find that the value that best fits the bulge's CMD is $(m-M)_{0}=29.04$. Nevertheless, the SFH was substantially unchanged within the uncertainties. For this reason, we retain \cite{Tully2013}'s value.\par

The synthetic CMD method not only enables us to determine the SFH of a stellar system, but also its total stellar mass, which represents a lower limit to the total dynamical mass of the system (i.e. baryonic matter plus dark matter halo). For the bulge of NGC 5474 we found a total stellar mass value of $(5.0 \pm 0.3) \times 10^{8} \,M_{\odot}$. According to hydrodynamical $N$-body simulations \citep{Pascale2021}, the hypothesis that the bulge is located within the galaxy's disc plane can be considered plausible provided that its total dynamical mass does not exceed $\sim 10^{8} \,M_{\odot}$. Indeed, a system more massive than that would: i) shift the entire galaxy's gravitational potential minimum, making the kinematic center of the disc coincide with the center of the bulge; ii) induce strong distortions in the HI velocity field map, inconsistent with observations \citep{Rownd1994,Kornreich2000}. Thus our result for the stellar mass seems to suggest that the putative bulge is not a real bulge orbiting within the plane of the disc, but rather an independent system moving on a polar orbit around NGC 5474.  

\subsection{SW over-density}
\begin{figure*}
	\centering
	\includegraphics[width=\textwidth]{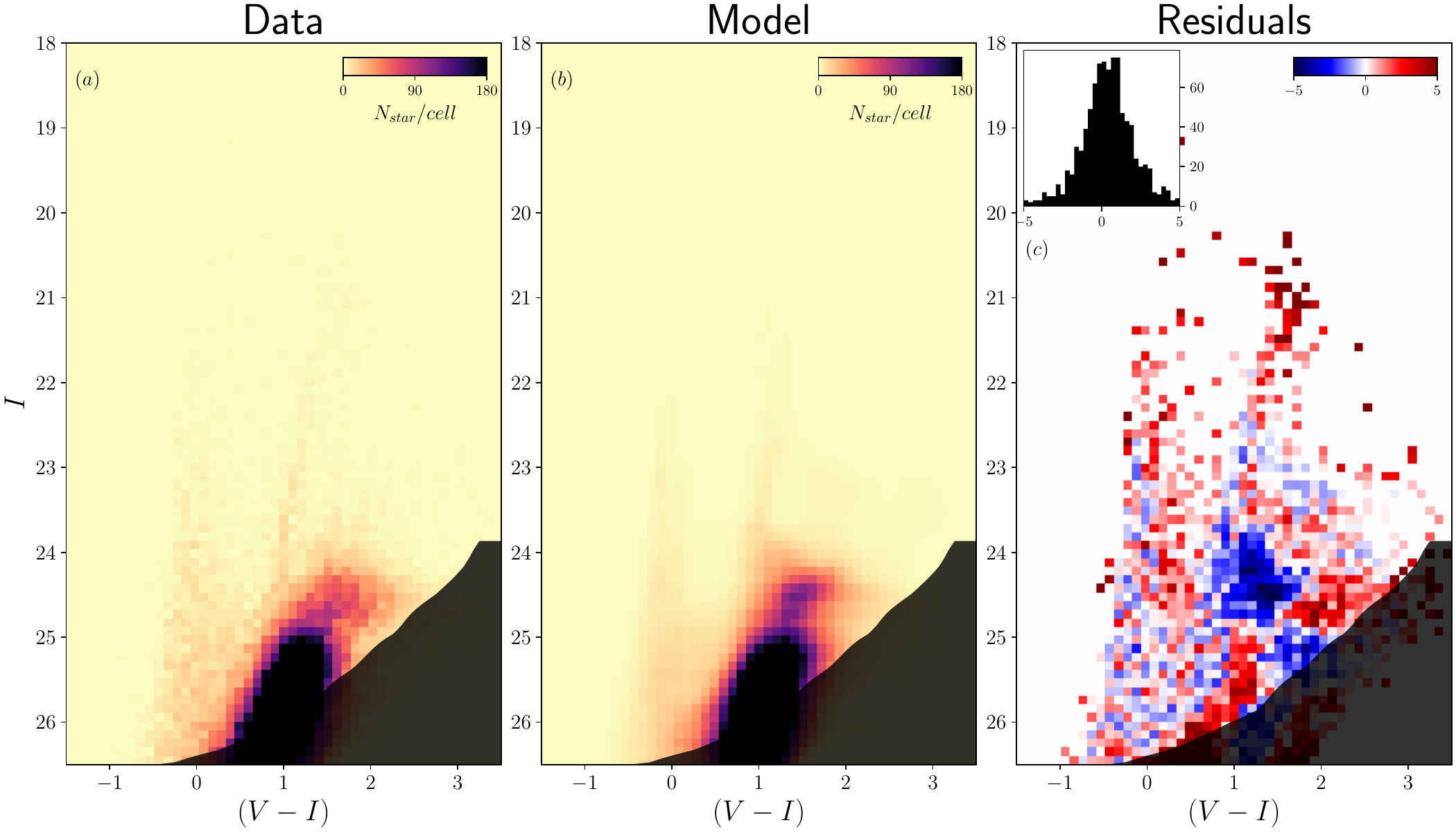}
	\includegraphics[width=0.49\textwidth]{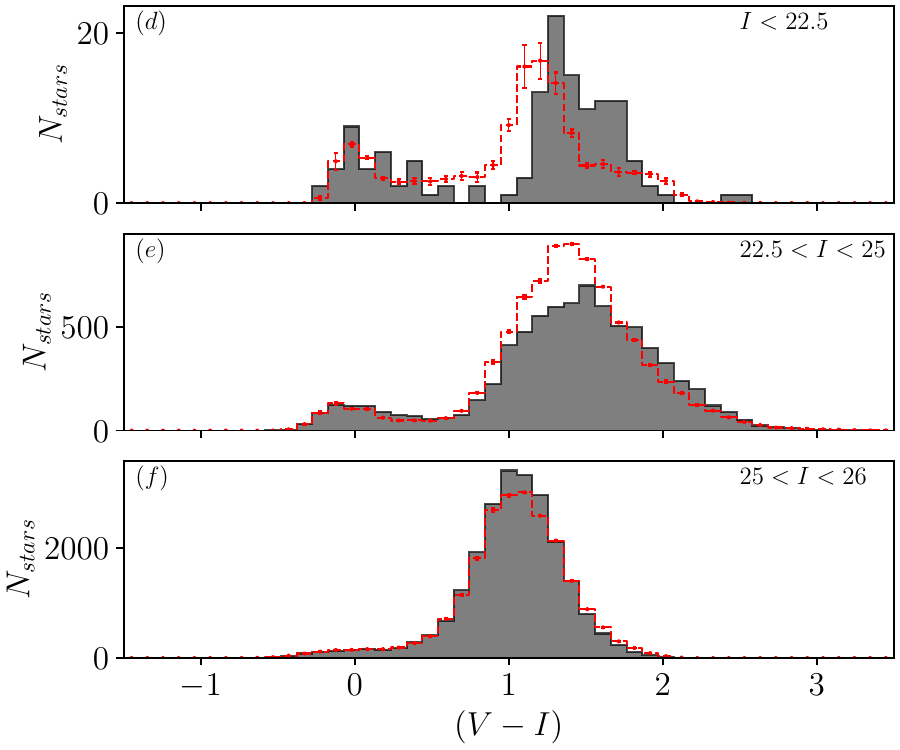}
	\includegraphics[width=0.49\textwidth]{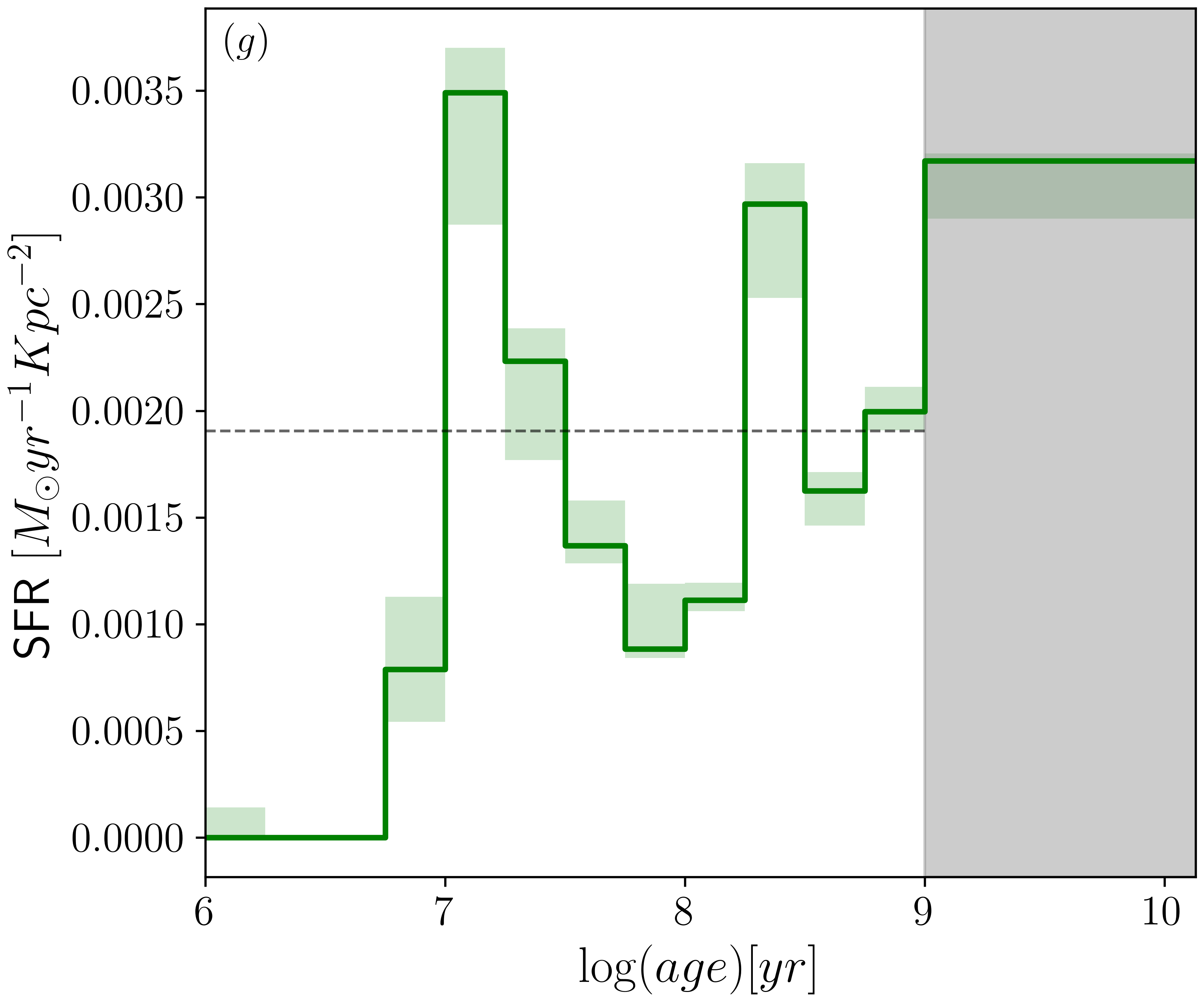}
	\caption{Same as Figure \ref{fig:bulge_bestmodel}, but for the South-West over-density region of NGC 5474. \label{fig:SWoverdensity_bestmodel}}
\end{figure*}
Figure \ref{fig:SWoverdensity_bestmodel} shows the results of the SFH analysis for the SW over-density region of NGC 5474. As for the bulge, we excluded from the fit the region of the Hess diagram where the completeness drops below $50\%$ (marked by the shaded area in panels (a)--(c)), to avoid being unduly affected by observational uncertainties.\par
The recovered SFH of the SW over-density (panel (g)), shows substantial activity between 1 and 14 Gyr ago, almost 1.5 times higher than the mean activity in the last Gyr (gray-dashed line). This result confirms that, like the putative bulge, also the SW over-density is dominated by a stellar population older then 1 Gyr, and possibly as old as 14 Gyr. However, unlike the putative bulge, the SW over-density shows two prominent SF peaks, around $10$ and $100$ Myr respectively.\par

The colour distribution in panel (d) of Figure \ref{fig:SWoverdensity_bestmodel} shows some differences between the data and the best fit model. In particular, SFERA 2.0 predicts again a slightly bluer red plume at $0.8 \lesssim (V-I)\lesssim 1.2$ and $I < 22$. Possible explanations for the observed discrepancy are: i) the effects of differential reddening. Indeed, even if NGC 5474 is affected by a low foreground extinction on average, close to the main spots of SF (mostly along the spiral arms) the internal extinction could be significant and very difficult to model; ii) massive stars in the brightest post-MS phase could be not properly modeled by the stellar evolution models. Indeed, massive young stars are subject to stochastic processes like strong stellar winds and mass losses \cite[see e.g.,][]{Chiosi1986,Puls2008}, that can significantly alter the magnitude and colour of the star. Again, in panel (e), our model over predicts the number of stars in the CMD region populated by AGB and TP-AGB. This discrepancy is probably due to the challenges of modelling such a complicated evolutionary phase \citep{Marigo2007,Marigo2017}, in which many complex phenomena occur (e.g., mixing of the convective layers, hot-bottom burning) which are not yet well understood in detail, and this leaves considerable uncertainty in the model predictions. Moreover, thermally pulsating AGB stars experience huge mass loss due to stellar winds \citep{Blocker1995}, which is still very hard to model and quantify. Nonetheless, the main features of the observed CMD are well reproduced by the model, within the uncertainties provided by the bootstrap approach (see panels (e)--(f)), and $90\%$ of the normalised residuals (panel (c)) are lower than $5 \sigma$.
Once the distance was allowed to vary, as explained in Section \ref{Bulge}, the value of distance modulus with the highest likelihood is $(m-M)_{0}=29.06$. Again, the corresponding SFHs stay largely within the uncertainties estimated through the bootstrap approach.
Finally, for the SW over-density of NGC 5474 we estimated a total stellar mass of $(5.2\pm0.1) \times 10^{8} \,M_{\odot}$.

\subsection{Disc}
\begin{figure*}
	\centering
	\includegraphics[width=\textwidth]{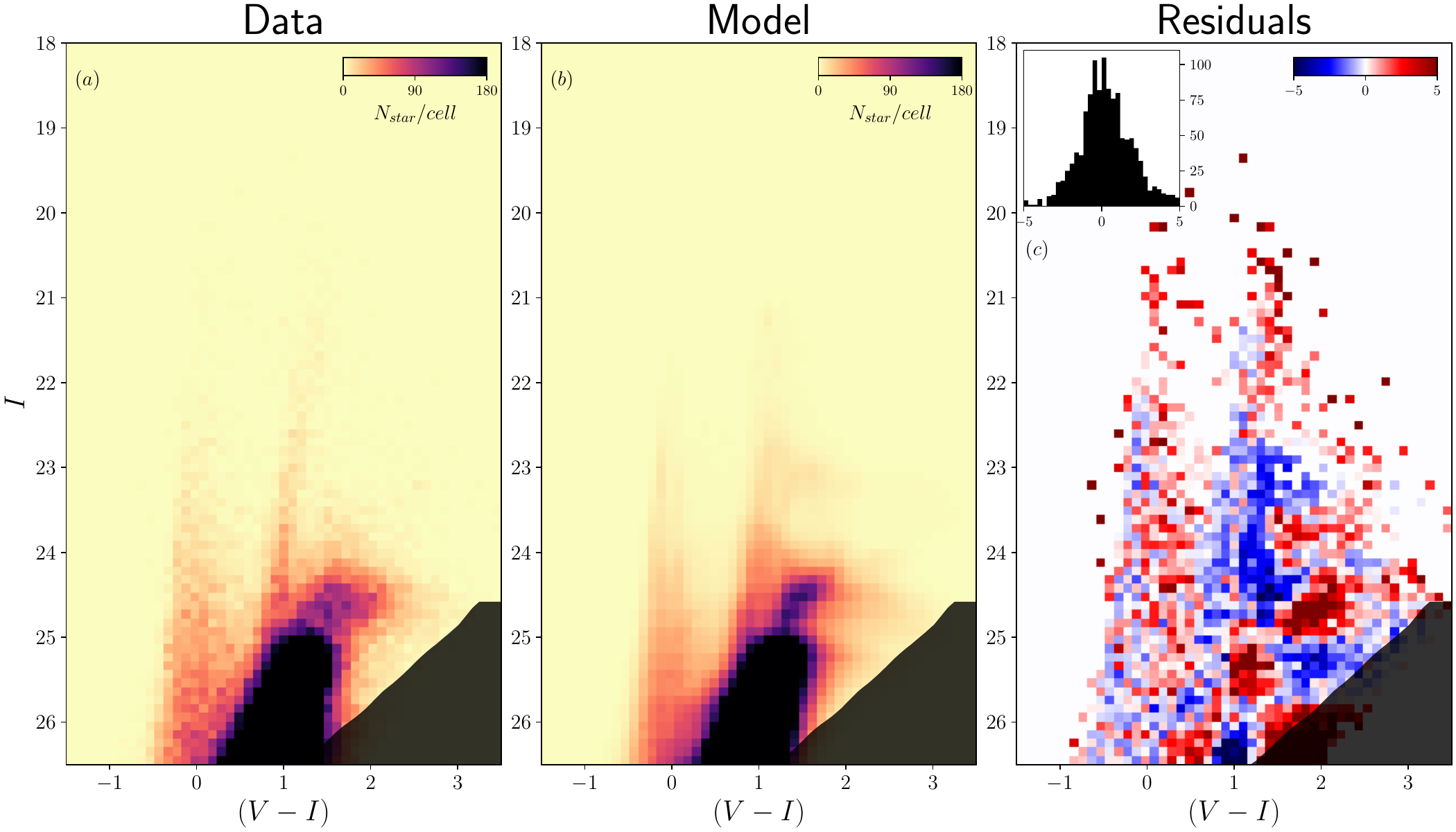}
	\includegraphics[width=0.49\textwidth]{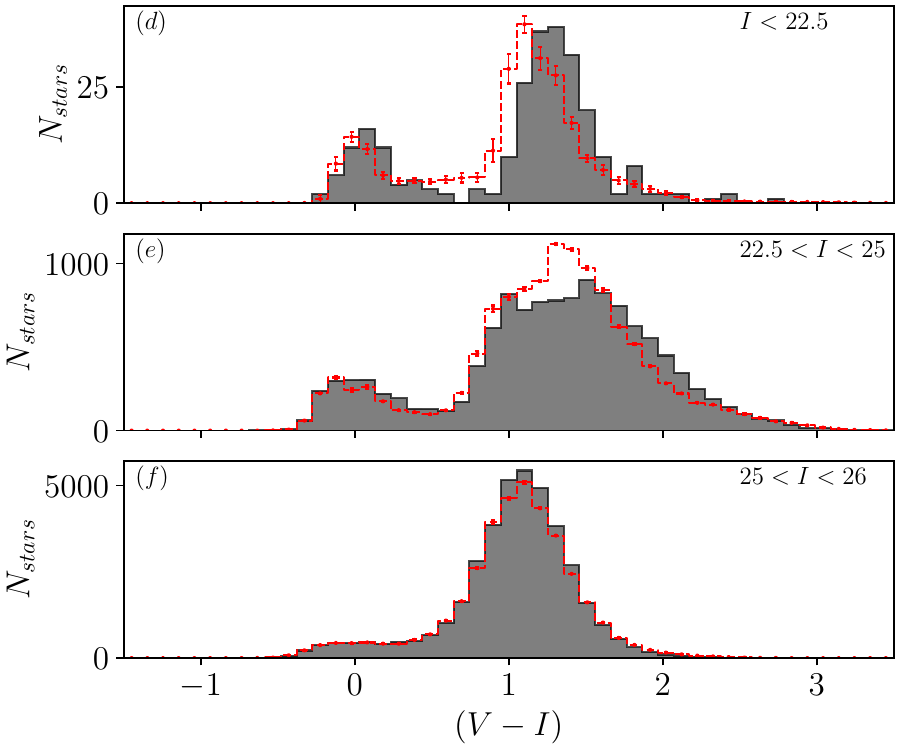}
	\includegraphics[width=0.49\textwidth]{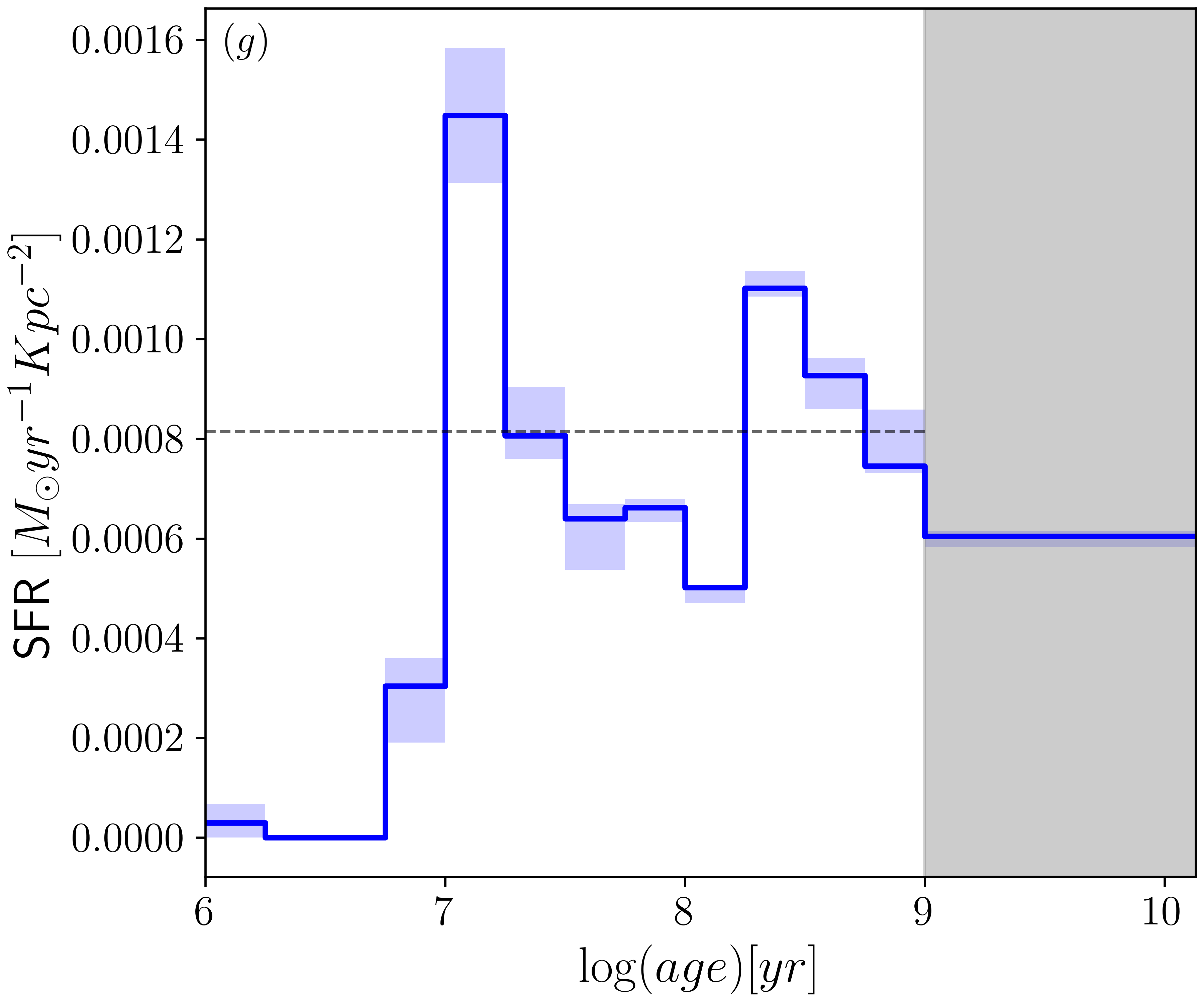}
	\caption{Same as Figure \ref{fig:bulge_bestmodel}, but for the stellar disc of NGC 5474. \label{fig:disc_bestmodel}}
\end{figure*}
As suggested by the presence of stellar populations of all ages in the CMD, the SF of the stellar disc of NGC 5474 has been rather continuous over the entire lifetime of the galaxy, without significant periods of enhanced SF activity (see panel (g) of Figure \ref{fig:disc_bestmodel}). In particular, we find again two peaks around $10$ and $100$ Myr, but the mean SFR in the last Gyr is higher or comparable to the activity in the oldest bin. Moreover, the duration of the peaks is similar to that of the dips. This behavior is consistent with a \textquotedblleft gasping\textquotedblright\ regime \cite[see e.g.,][]{Ferraro1989,Tosi1991,Marconi1995,Gallart1996,Tolstoy1996,Weisz2012,Annibali2022}, rather than a bursty one (short episodes of strong SF activity separated by long quiescent phases).\par

In panels (a)--(c) of Figure \ref{fig:disc_bestmodel} we show the observed Hess diagram, the best-fit model obtained with SFERA 2.0, and the normalised residuals, respectively. Again, we excluded from the fit the cells of the CMD with an estimated completeness value below $50\%$ (marked by the shaded area in the Hess diagrams). The residuals show some systematic trends, especially in the region immediately above the RGB-tip ($1\lesssim(V-I)\lesssim 2.5$ and $22\lesssim I\lesssim 25$). This region is mostly composed of AGB and TP-AGB stars, for which we mentioned in the previous section are quite challenging to model. The best fit model seems to under-predict again the number of stars in the brightest part of the red plume ($(V-I)\sim 1$ and $I<22.5$). Nevertheless, the upper MS phase appears well reproduced, together with the RGB region. We also point out that, again, $90\%$ of the residuals are below the $5\sigma$ limit.\par
In terms of colour distribution, our best model seems to reproduce quite well the observations. Despite this, the best model colour distribution displayed in panel (d) of Figure \ref{fig:disc_bestmodel} shows the same problems in reproducing the red plume and the number of AGB stars encountered in the previous sections.\par
Same as for the bulge and the SW over-density regions, we left the distance free to vary in the range $(m-M)_{0} = [29.02,29.42]$. The resulting best-fitting distance came out to be $(m-M)_{0}=29.04$. Once again, the recovered SFHs are consistent with the values derived using the fixed distance modulus from \cite{Tully2013}. Finally, For the NGC 5474's disc we found a total stellar mass of $(5.3\pm0.1) \times 10^{8} \,M_{\odot}$.

\subsection{Bulge vs. SW over-density vs. disc}
\begin{figure}
	\centering
	\includegraphics[width=\columnwidth]{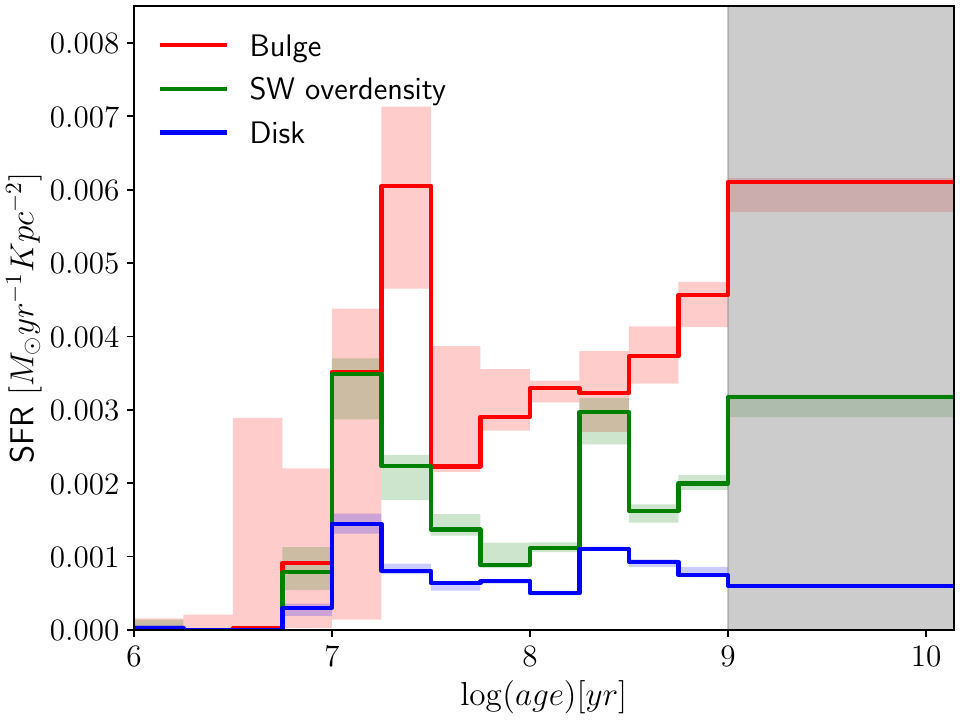} 
	\caption{Comparison between the SFHs of the bulge (red line), SW overdensity (green line), and disc (blue line) of NGC 5474. All three SFHs are normalised to the area of their respective region. The epochs older than 1 Gyr are shaded gray due to the lack of good age indicators in the CMD.\label{fig:SFHs_comparison}}.
\end{figure}

A direct comparison between the SFHs of the three selected regions in NGC 5474 (i.e., bulge, SW over-density, and disc) is shown in Figure \ref{fig:SFHs_comparison}. In the last Gyr, the SFHs of SW over-density (in green) and the disc (in blue) show quite similar behaviors in terms of overall shape, both showing two main peaks of SF around $\sim 10$ Myr and $\sim 100$ Myr ago, but not in terms of overall rate, since the average activity of the disc is significantly lower. The main difference between the two regions appears in the oldest bin (1--14 Gyr ago), where the SW over-density displays a pronounced activity, while the disc has been relatively inactive. On the other hand, the bulge (red line) is remarkably different, showing a strong SF activity older than 1 Gyr, to which follows a steady decrease in activity, interrupted only by a very uncertain peak around 20-35 Myr ago.\par 

It is worth noting that the peak of SF around $100$ Myr experienced by the SW over-density and the disc roughly coincides with the close passage of NGC 5474 within M101's outer disc predicted by \cite{Linden2022} simulations, while the bulge shows no significant enhancement in its SF activity in the same time window. Even more interestingly, all three regions seems to have experienced a synchronized burst of SF around $10-35$ Myr ago. These SFHs might be explained by a first interaction between M101 and NGC 5474, capable of driving the burst around $100$ Myr, while a more recent interaction between NGC 5474 and the putative bulge could explain the youngest burst of SF activity present in all three component. Oddly, all three regions seem to have ceased SF simultaneously in the last $10$ Myr.\par

As discussed at the beginning of this section, the AMRs of the three regions are not well constrained due to the absence in the CMDs of a feature able to break the age-metallicity relation (i.e., MS turn-off). Nevertheless, we were able to infer an average metallicity values for the three selected regions. Unfortunately, the uncertainties do not allow us to discern possible significant differences in the metal content of the three components. However, the presence or absence of differences in metallicity content between the three components, though an important piece of information, would not automatically prove or imply that the `bulge' is a distinct object.

\section{Summary and conclusions}\label{Discussion and conclusions}

This paper was devoted to the analysis of the resolved stellar populations and SFH of the peculiar spiral dwarf galaxy NGC 5474. Following the works of \cite{Bellazzini2020} and \cite{Pascale2021} who renewed the interest in this peculiar galaxy, we presented for the first time an extensive analysis of the HST LEGUS resolved-star photometry, based on a quantitative treatment of the incompleteness of the data and of the star formation histories in different galaxy regions using SFERA 2.0. The observed CMD of NGC 5474 (Figure \ref{fig:CMD} and \ref{fig:CMD_plus_iso}) shows prominent blue and red plumes, likely populated by young upper-MS and BL stars ($\lesssim 100$ Myr). Moreover, it displays a well-defined RGB and AGB red-tail, composed by old ($\gtrsim 1$ Gyr) and intermediate ($0.1\lesssim$ age $\lesssim 4$ Gyr) age stars, respectively.\par

Concerning the spatial distribution, we found that stars younger than $\sim 100$ Myr show a wide spiral pattern reaching the center of the bulge, where they are arranged in a bar-like structure. Moreover, the recent activity is mainly on the South-West portion of the galaxy's disc (Figure \ref{fig:MS_BL_spatial_distr}).\par

The AGB and RGB stars spatial distributions (Figure \ref{fig:AGB_RGB_spatial_distr}) show two dominant structures, the bulge and a South-West over-density, separated by a local paucity of stars (gap), which coincides with the dynamical center of the HI distribution. As corroborated by extensive artificial star tests, the gap is real.\par

To study the SFH of NGC 5474, we identified three different regions within the galaxy (i.e., the bulge, the South-West over-density, and the disc). Our main findings are:
\begin{itemize}
	\item The putative bulge SFH shows a strong activity older than $1$ Gyr (possibly as old as 14 Gyr), but also clear evidence of prolonged SF activity up to at least 10 Myr ago, with a very uncertain peak around $\sim 20-35$ Myr, a rather unusual feature in \textquotedblleft classical\textquotedblright\ bulges. Moreover, we provided a lower limit to the total bulge's mass of $(5.0 \pm 1.0) \times 10^{8} \,M_{\odot}$. This value is significantly higher than the limit found by \cite{Pascale2021} for a compact system to be orbiting within the galaxy's disc plane without inducing strong distortions in the HI velocity field. Thus, both the large mass and the presence of young SF seem to support the hypothesis of the bulge as a compact early-type dwarf galaxy orbiting around NGC 5474. In this case, the observed off-set between the bulge and the center of the HI distribution could be easily explained by projection effects, as explained in \cite{Pascale2021}.
	\item The SW over-density's SFH shows three prominent peaks of SF, one at epochs earlier than 1 Gyr ago, $\sim 1.5$ times higher than the average SF activity in the last Gyr, validating the ancient nature of this population, and two younger peaks at $\sim 10$ and $\sim 100$ Myr, respectively. The peak around $\sim 100$ Myr is broadly consistent with a possible interaction between NGC 5474 and M101. 
	\item  In the last Gyr, the disc of NGC 5474 shows a rather similar SFHs in terms of overall shape and trend with respect to the SW over-density, but not in overall rate. Indeed, the disc average activity is consistently lower, and more `flat' than in the SW over-density. Moreover, the disc shows a significantly lower rate of SF at age $> 1$ Gyr with respect to the putative bulge and SW over-density. Its behavior is consistent to what in the literature is called a `gasping' regime.
    \item For NGC 5474, we found a total stellar mass of $(1.55\pm0.03)\times 10^{9} M_{\odot}$, within a radius of $\sim 3$ kpc. This value falls near the dynamical mass range reported by \cite{Rownd1994} for the galaxy within a radius of $\sim 5$ kpc ($2.0-6.5 \times 10^{9} \, M_{\odot}$). This result may suggest that NGC 5474 is not a dark matter dominated system \citep{Moreno2022}.   
\end{itemize} 

In conclusion, while this study is not sufficient to ascertain once and for all the real nature of the putative bulge and of the many peculiarities of NGC 5474, it puts forward new important information that seems to strongly support the idea of the putative bulge being and `independent' dwarf galaxy orbiting around the disc of NGC 5474. Moreover, it provides for the first time a detailed SFH analysis of its major stellar components, an important piece of information that was lacking from the overall picture of this highly disturbed, but fascinating galaxy. These SFHs will be also crucial for a possible future chemo-dynamical modeling of the galaxy.

\section*{Acknowledgements}
This work is largely based on the Master Thesis of GB supervised by MC, discussed in the 2020-2021 academic year at Pisa University, and then completed at Stockholm University.
MC acknowledges the INFN (Iniziativa specifica TAsP).
AA acknowledges support from the Swedish Research Council, Vetenskapsr{\aa}det (2021-05559, 2016-05199)
K.G. is supported by the Australian Research Council through the Discovery Early Career Researcher Award (DECRA) Fellowship (project number DE220100766) funded by the Australian Government. 
K.G. is supported by the Australian Research Council Centre of Excellence for All Sky Astrophysics in 3 Dimensions (ASTRO~3D), through project number CE170100013. The authors thank the anonymous reviewer for their careful reading and constructive feedback.

\section*{Data Availability}
The data underlying this article are associated with the HST Treasury program 13364 (PI: D. Calzetti), and available on the LEGUS survey public access table on MAST archive, at https://archive.stsci.edu/prepds/legus/dataproducts-public.html.

\appendix
\section{SFERA 2.0}\label{app:SFERA2.0}
In this appendix we present in detail the updated synthetic CMD generation algorithm, SFERA 2.0. The accuracy and reliability of the code are tested with mock galaxies with different known SFHs and metallicity distributions.\par
In our methodology, the generation of synthetic CMDs consists of a two-step process: (i) producing 'clean' CMDs and (ii) convolving them with the distance, reddening, and photometric errors specific to the target galaxy under investigation. We divide this process into two steps because creating 'clean' CMDs is the most computational-expensive part of the method, but it is a one-time requirement, whereas the incorporation of observational errors depends on the particular target under study \citep{Dolphin2002}.\par

\subsection{Adaptive interpolation scheme}
In the previous version of the code (SFERA, \citealt{Cignoni2015,Sacchi2018}), the construction of a basis CMD involved extracting millions of synthetic stars randomly drawn from a uniform IMF, subsequently re-scaled to a physical IMF \citep[i.e.,][]{Salpeter1955,Kroupa2001}. Although Monte Carlo routines are attractive for their simplicity, they can not sample properly all the stellar evolutionary features present in any galaxy CMD. Indeed, the density of points on the CMD varies by many orders of magnitude between the MS, the most populated evolutionary stage where stars burn H in the core, and later evolutionary phases (i.e., RGB, BL, AGB). This is due to both the IMF, generally heavily weighted towards lower mass stars \citep{Kroupa2001} (a difficulty already addressed by SFERA), and to the fact that stellar lifetimes in different evolutionary phases differ by many orders of magnitudes (e.g., a $15$\MSUN star burns hydrogen in the core for $\sim 11$ Myr, while it burns carbon for $\sim 3000$ yr). Instead, our new algorithm aims to produce a `true' model synthetic CMD, effectively sampling all possible combinations of masses, ages, and metallicities comprising the basis CMD.\par
\begin{figure}
	\centering
	\includegraphics[width=\columnwidth]{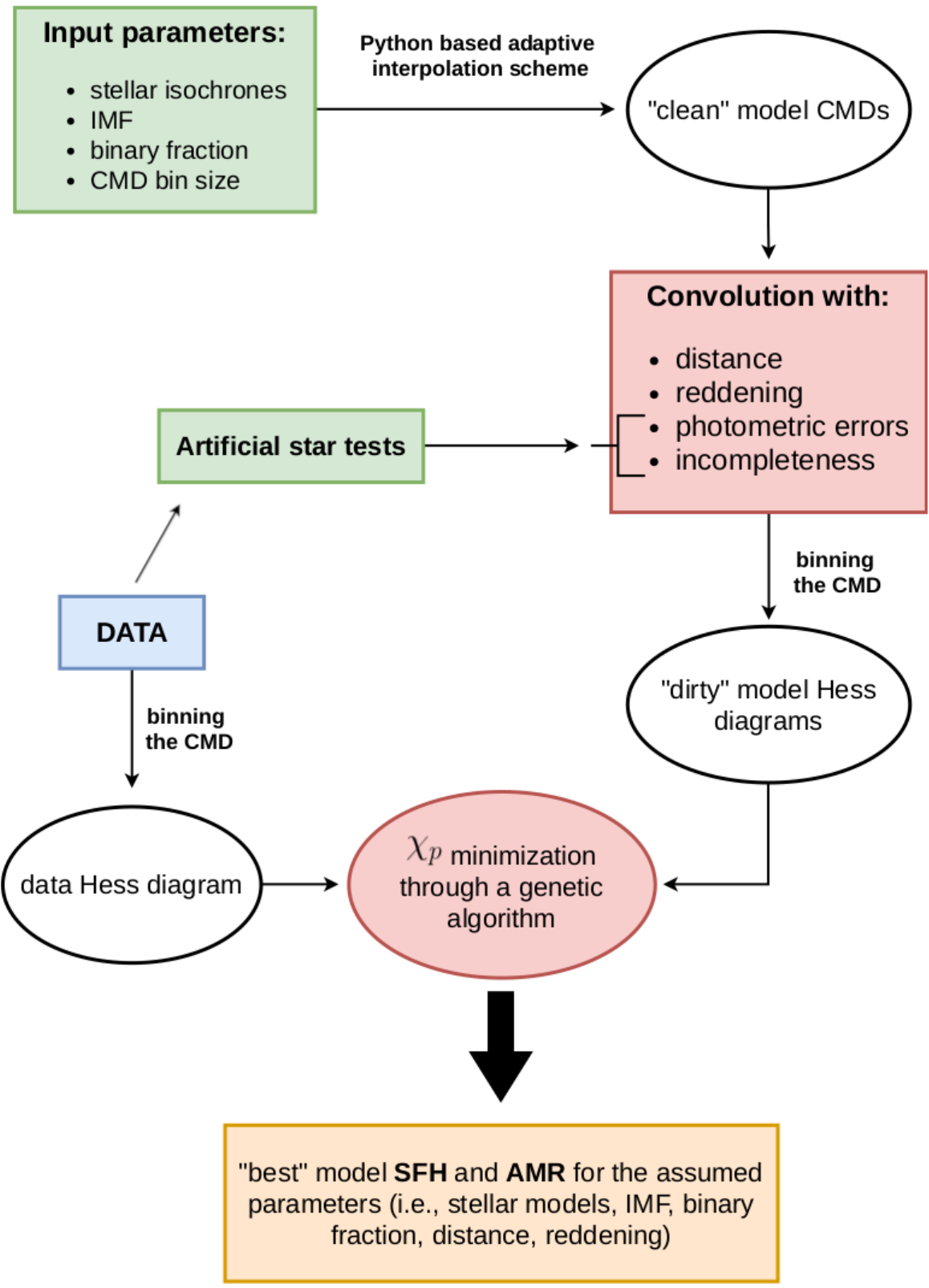} 
	\caption{Schematic view of SFERA 2.0, from the computation of the synthetic CMDs to the minimization of the residuals with the data.\label{SFERA_2.0_diagram}}
\end{figure}

As shown in the scheme in Figure \ref{SFERA_2.0_diagram}, to generate a partial synthetic CMD, the code requires the following input parameters: a high resolution library of stellar isochrones, an IMF, an assumed binary fraction, and a colour-magnitude grid size into which the synthetic CMD and the observed one will be binned for the comparison. The choice of CMD bin size is a crucial point. Ideally, one would like to divide the CMD with the finest colour-magnitude grid possible, in order to  reach the maximum time resolution possible. Practically, one needs to consider that the smallest photometric errors affecting distant galaxies are generally of the order of $0.01$ mag. Moreover, most stellar evolutionary models are provided with a resolution lower than $0.01$ mag. Finally, the computational limitations must be taken into account too. A too small bin size choice would make the extremization of the likelihood very time-consuming. Thus, the most reasonable choice of CMD bin size generally lies between $0.05$ and $0.1$, both in colour and magnitude.

The first step to generate a `true' model CMD is to calculate a sufficiently large number of isochrones, so that the maximum separation between adjacent isochrones in the CMD is smaller than the selected CMD bin size. Due to the fact that stellar isochrones are generally provided with a coarse mass step, the second step is to perform a mass interpolation along each isochrone. To this purpose, the isochrones of different ages are considered one at a time. The various stellar evolutionary phases of each isochrone are then interpolated in such a way that the maximum distance between two adjacent stars is always smaller than the selected CMD bin size. This adaptive interpolation scheme allows us to sample \textit{uniformly} all the different stellar evolutionary phases (from the pre-MS to the TP-AGB), even if characterized by different evolutionary timescales and stellar masses. This new interpolation scheme is able to populate all CMD bins along the isochrone with a fixed number of points, unlike the previous Monte Carlo algorithm, in which one can only choose the total number of synthetic stars to be generated. At this point, the synthetic stars are uniformly distributed in the CMD, but unevenly distributed in mass, not following a specific IMF. To account for this issue, each interpolated point along the isochrone is weighted according to its mass and the assumed IMF. Thus, the new model points do not represent a single star anymore, but rather a `probability' density distribution in the CMD.\par

At this point, synthetic stars with lifetimes shorter than the age of the isochrone are considered `dead', thus not visible in the CMD, but taken into account in the total mass extracted. The procedure outlined above is then repeated for each isochrone.\par
In the previous Monte Carlo approach, accounting for the population of unresolved binary systems was straightforward. A given percentage of primary stars was coupled with a companion, whose mass was a random fraction of the primary. In the framework of the new adaptive interpolation scheme the treatment of binary systems becomes more complex. Indeed, as with the primary stars, it is now necessary to sample the range of all possible secondary star masses to at least the accuracy of the CMD bin size. In order to accomplish this, for each primary star selected to be in a binary system, multiple secondary stars are extracted with masses uniformly sampled in the interval $[0.2 \, M_{\odot},M_{\rm primary}]$. The primary is then coupled, in turn, with each one of the extracted secondaries. The position in the CMD of the unresolved binary system is given by the sum of the primary and secondary star fluxes, while the weight of the system in the CMD is given by the weight of the primary star, divided by the number of secondary stars extracted. \par
The final result of this new model generation process can be thought of as a `probability' density distribution of stars in the CMD, whereas the previous Monte Carlo models were just stochastic representations. The crucial point is that all possible masses, ages, and binary combinations are accounted for in these new synthetic CMDs, and that regardless of the selected CMD bin size, every evolutionary feature is properly sampled. Conversely, low-density regions in the model CMDs generated by a Monte Carlo algorithm, are poorly handled by the likelihood extremization process, reducing the SFH recovering power of the synthetic CMD method. From this point of view, SFERA 2.0 is now more robust.

\subsection{Convolution with observational conditions}
Observed CMDs are never simple samplings of model isochrones. Distance of the target galaxy, foreground reddening due to the presence of dust along the line of sight, incompleteness, and photometric errors all considerably complicate the matter. In order to generate an accurate model CMD, all of these factors must be carefully taken into account. First, each point of the `clean' model CMD is placed at the distance of the target galaxy (in order to transform absolute to apparent magnitudes), and corrected for the assumed foreground extinction. Completeness and photometric errors of the data are evaluated performing artificial star tests on the real images. For each bin of the CMD, the completeness level is provided by the ratio between injected and recovered artificial stars, while the distribution of photometric errors is given by the difference between their output and input magnitude values. 

\begin{figure}
	\centering
	\includegraphics[width=\columnwidth]{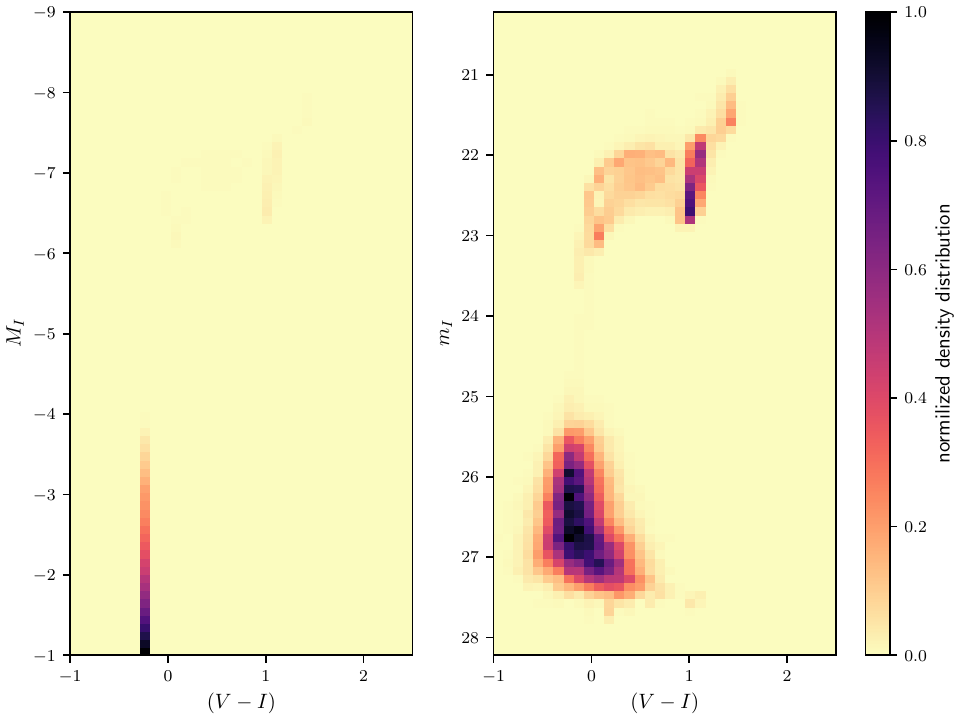}
	\caption{Comparison between two model Hess diagrams with a constant SFR between 20-30 Myr, uniform metallicity between 0.003 and 0.004, a Kroupa IMF, and a $30\%$ of unresolved binary systems, before and after the convolution with photometric errors and incompleteness typical of HST for a galaxy at the distance of NGC 5474. Both partial CMDs are binned with a bin size of 0.1 both in colour and magnitude. The colour map is proportional to the normalised stellar density.}
	\label{cleanvddirtyhess}
\end{figure}

Finally, the `dirty' partial CMDs are converted into Hess diagrams (mainly for an easier storage), where the number density of stars expected in each CMD bin is given by the sum of the weights of all model points that fall within that bin. In Figure \ref{cleanvddirtyhess} we show a comparison between a `clean' model Hess diagram (left panel), and the one obtained after the convolution with photometric errors and incompleteness typical of HST photometry for a galaxy at $\sim 7$ Mpc (right panel). This CMD was generated for ages between 20-30 Myr, a uniform metallicity in the interval $0.003-0.004$, a \cite{Kroupa2001} IMF, and a $30\%$ of unresolved binary systems.

\subsection{Models vs. data} 
The extremization of the likelihood is implemented in SFERA 2.0 following a Poissonian statistics, looking for the linear combination of partial Hess diagrams that minimizes the likelihood between models and data \citep{Dolphin2002}. As already discussed in Section \ref{SFH_recovery_method}, the high number of free parameters, together with the complex nature of the parameters space that needs to be explored, requires the implementation of a global minimization algorithm. In particular, SFERA 2.0 implements a classical genetic algorithm-based minimization routine. This allows to take advantage of the exploration ability of the former, which is able to scan the parameters space in more points simultaneously, with negligible memory of the initial conditions \citep{Charbonneau1995}. From the minimization process we obtain the SFH and AMR that better reproduce the data for the assumed stellar models, IMF, binary fraction, distance and reddening.\par
Finally, to estimate the  statistical uncertainties around the `best' model, we perform a bootstrap test. The original data are randomly resampled with replacements, producing a set of pseudo-data. This mimics the observational process: if the observational data are representative of the underlying distribution, the data produced with replacements are copies of the original one with local crowding or sparseness. The SFH recovery algorithm is performed on each of these replicated datasets, and the result is a set of  `best' parameters. The confidence interval is then defined as the interval that contains the 5$^{\rm th}$ and 95$^{\rm th}$ percentiles of the parameter distribution.

\section{SFHs and metallicity distribution recovery tests}
In order to test the robustness of the new star-formation recovery code, we performed some tests by applying SFERA 2.0 to mock galaxies, generated with known SFHs and AMRs. The simulated data are convolved with the typical photometric errors and incompleteness of NGC 5474 (at a distance of $\sim 7$ Mpc from us), observed with the HST/ACS WFC. It is important to stress that galaxies at this distance are among the furthest ones for which we can recover the SFH for the entire Hubble time, given the resolution power of modern instruments. Indeed, the $50\%$ completeness limit at $\sim 26.5$ mag is located approximately $1$ mag below the oldest RGB-tip. All mock galaxies have been generated using a classical Monte Carlo algorithm, assuming the PARSEC-COLIBRI stellar models, a \cite{Kroupa2001} IMF, and a $30\%$ of unresolved binary systems.\par
\begin{figure*}
	\centering
	\includegraphics[width=0.85\textwidth]{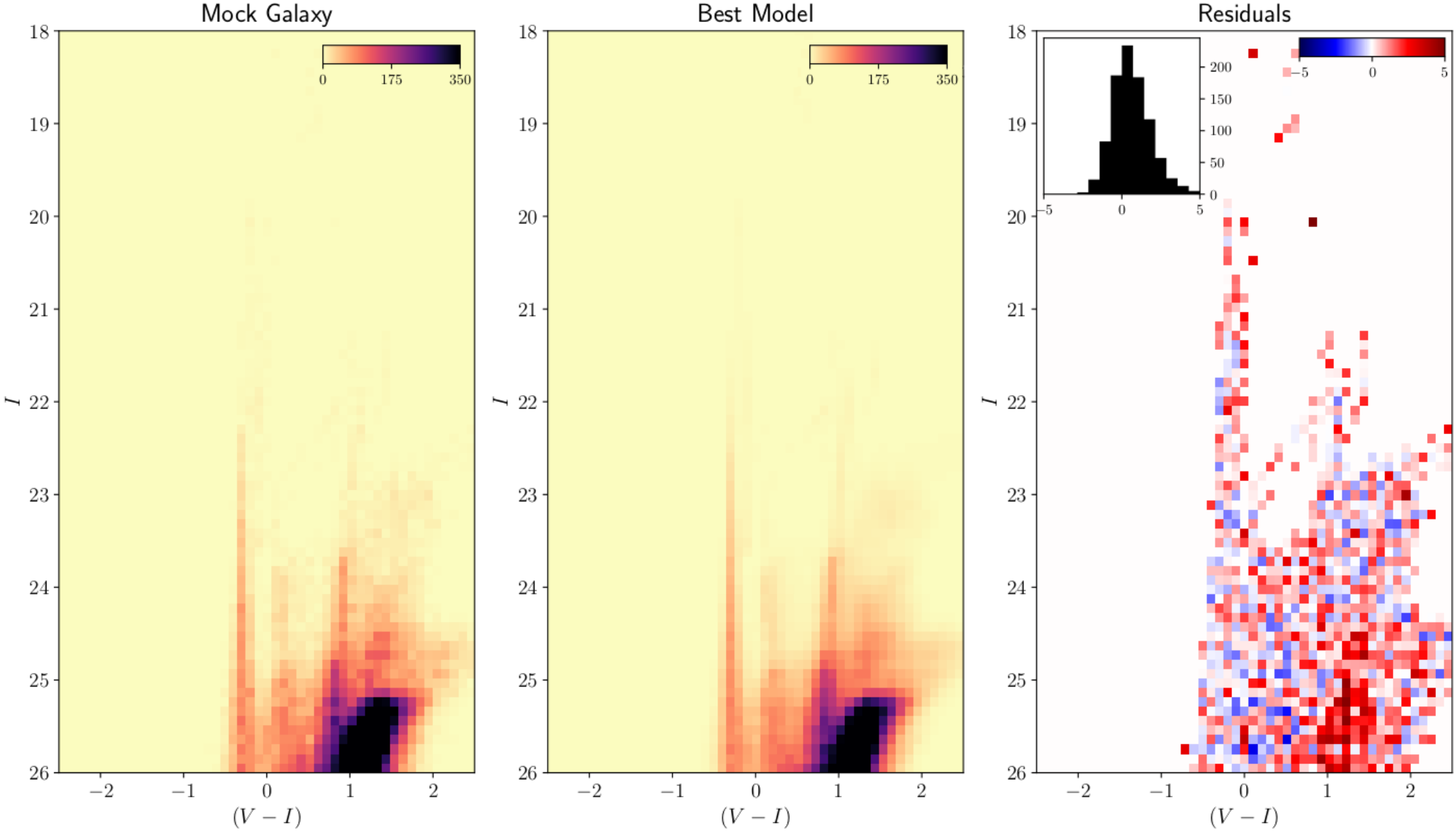}
	\includegraphics[width=0.85\textwidth]{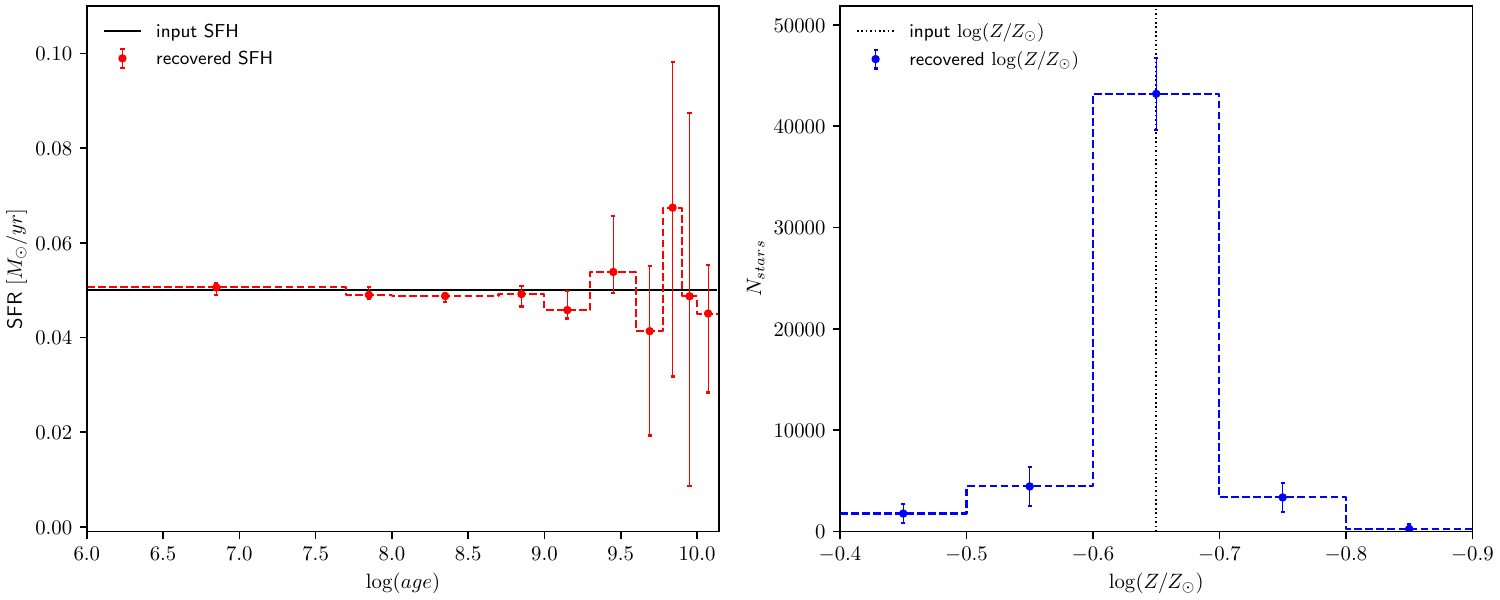}
	\caption{\textit{Top panels}. Mock galaxy (left panel), recovered (middle panel), and residuals (right panel) Hess diagrams. The residuals are normalised. The normalised residuals distribution is also shown in the inset panel. \textit{Bottom panels}. Recovered SFH and metallicity distribution with SFERA 2.0 for a mock galaxy generated with a constant SF from 13 Gyr to 1 Myr ago, and a uniform metallicity in the interval $\log(Z/Z_{\odot})=[-0.7,-0.6]$. \label{fig:sfhflat}}
\end{figure*}
\begin{figure*}
	\centering
	\includegraphics[width=0.85\textwidth]{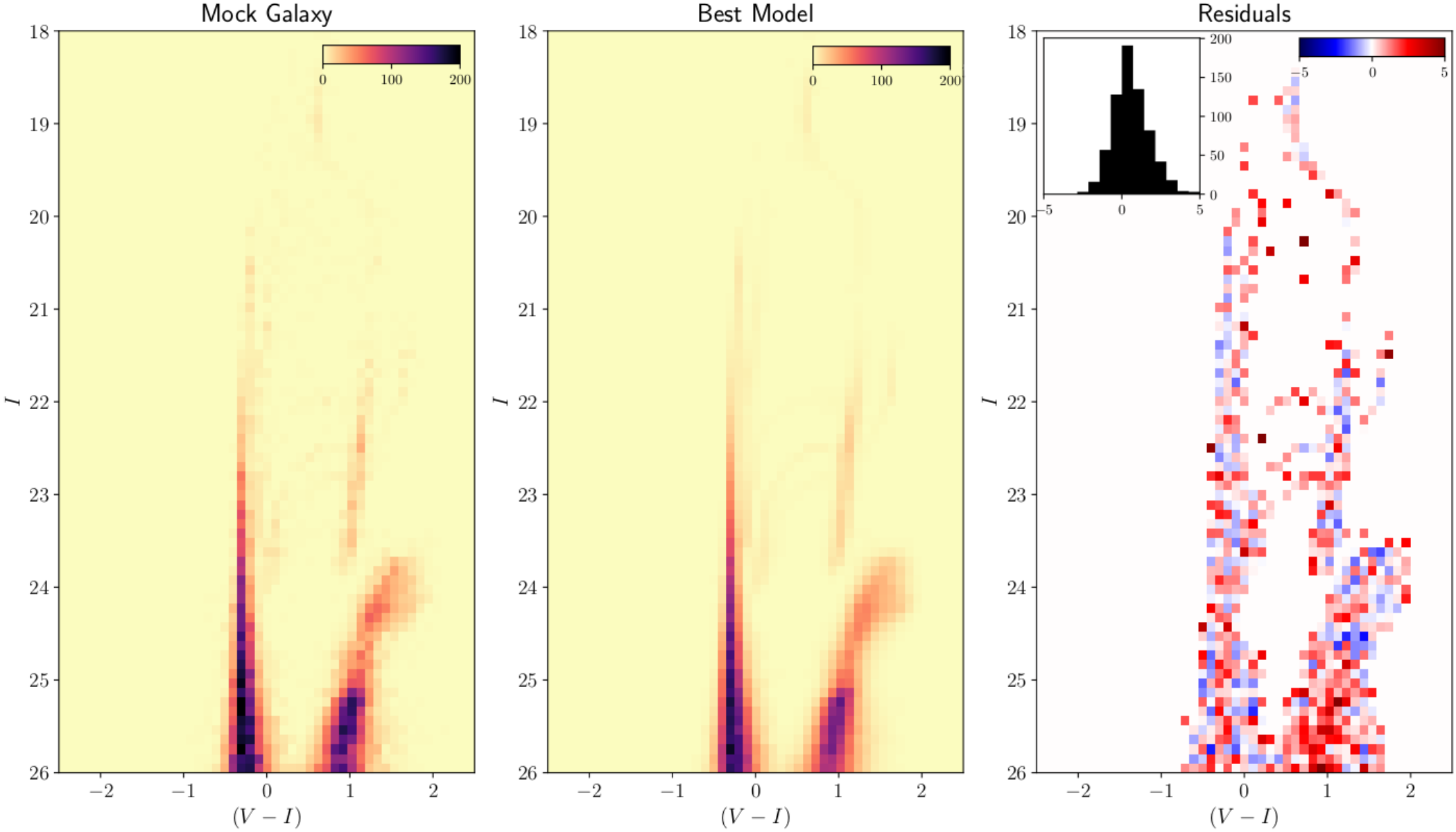}
	\includegraphics[width=0.85\textwidth]{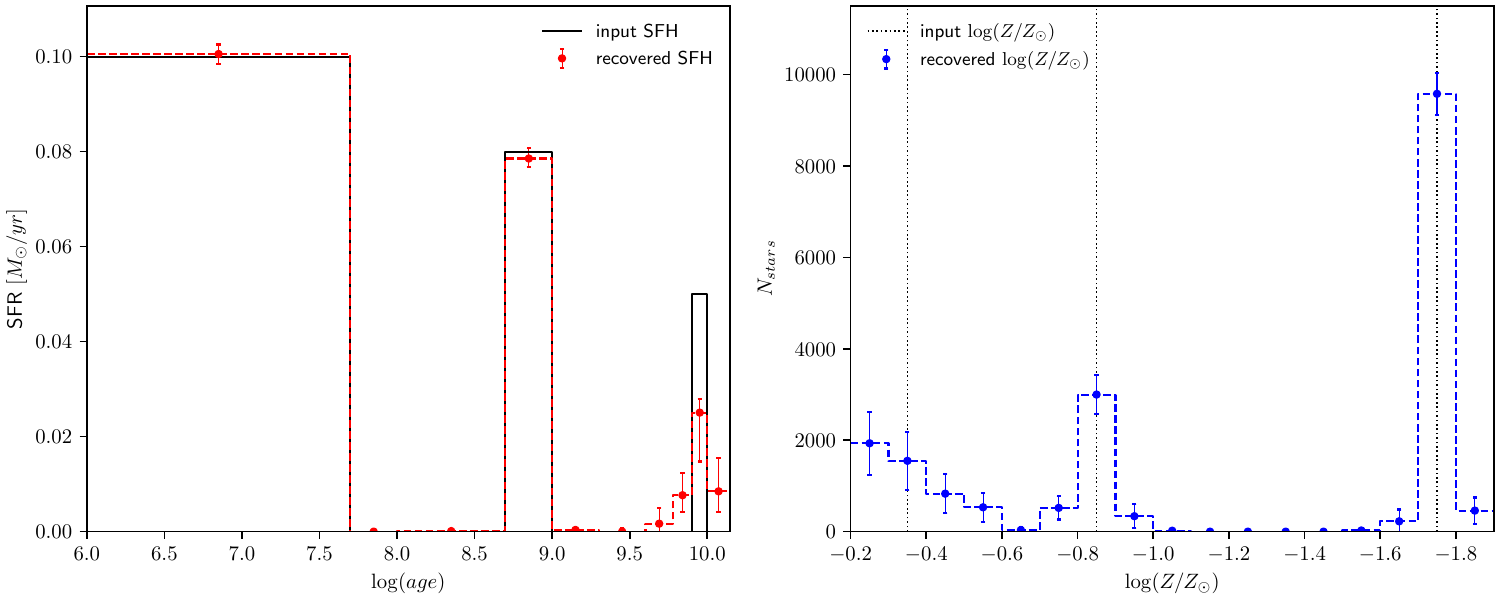}
	\caption{\textit{Top panels}. Mock galaxy (left panel), recovered (middle panel), and normalised residuals (right panel) Hess diagrams. The residuals distribution in also shown in the inset panel. \textit{Bottom panels}. Recovered SFH (left panel) and metallicity distribution (right panel) with SFERA 2.0 for a mock galaxy composed of three different episodes of SF; a recent one between $1-50$ Myr with a constant SFR of $0.1 \, M_{\odot}\, {\rm yr}^{-1}$ and a metallicity of $\log(Z/Z_{\odot}) = -0.35$, one of intermediate age (0.5--1 Gyr) with a SFR of $0.08 \, M_{\odot}\, {\rm yr}^{-1}$ and a metallicity of $\log(Z/Z_{\odot}) = -0.85$, and the oldest one between 8--10 Gyr, with a SFR of $0.05 \, M_{\odot}\, {\rm yr}^{-1}$ and $\log(Z/Z_{\odot}) = -1.75$.}
	\label{fig:sfhmetvar}
\end{figure*}
The first test was performed on a mock galaxy generated with constant SFR ($0.05 \, M_{\odot}\, {\rm yr}^{-1}$) from 13 Gyr to 1 Myr ago, and a uniform metallicity in the interval $\log(Z/Z_{\odot})=[-0.7,-0.6]$. The results are shown in Figure \ref{fig:sfhflat}. SFERA 2.0 seems to have recovered the input SFH and metallicity distribution quite accurately. The input SFH is always recovered within the uncertainties of the reconstruction, and the input metallicity is well recovered within 0.1 dex. Indeed, the observational Hess diagram is well-fitted by the best model, with $90\%$ of the residuals within $\pm 2 \sigma$ (see the inset panel). The only significant discrepancy is a large systematic feature in the region $1\lesssim(V-I)\lesssim$2 and $25\lesssim I\lesssim26$. This reflects the fact that the only signature of the oldest activity comes from old ($> 1$ Gyr) RGB stars, much more packed in the CMD than other stellar phases.\par 
The last test was performed on a mock galaxy composed of three episodes of SF with different ages, duration, and metallicity values. A very young burst between 1 and 50 Myr age, with a constant SFR of $0.1 \, M_{\odot}\, {\rm yr}^{-1}$ and a metallicity of $\log(Z/Z_{\odot}) = -0.35$, one between 0.5 and 1 Gyr ago, with a SFR of $0.08 \, M_{\odot}\, {\rm yr}^{-1}$ and a metallicity of $\log(Z/Z_{\odot}) = -0.85$, and the oldest one between 8 and 10 Gyr ago, with a SFR of $0.05 \, M_{\odot}\, {\rm yr}^{-1}$ and $\log(Z/Z_{\odot}) = -1.75$. The results are shown in Figure \ref{fig:sfhmetvar}. Also in this case the code recovers well the two youngest episodes of SF, while the oldest episode is recovered much broader than the input one, with some of the SFR leaked in the near time bins. This result strongly depends on the fact that, at the distance of NGC 5474, the CMD of the galaxy does not reach the ancient MS turn-off, and the SFH relies on poorer ages indicators (e.g., RGB, AGB and TPAGB). Thus, the resolution at the oldest epochs (age $\gtrsim 8$ Gyr) is no more than a few Gyr. We point out that limiting the SFH studies to galaxies within 1 Mpc, where the ancient MS turn-off can be easily detected, would prevent us from covering all possible galactic environments and dwarf galaxies morphological types that populate the Local Volume. Moreover, the recovered metallicity distribution for the two oldest SF episodes is in excellent agreement ($\pm0.1$ dex) with the input metallicity. On the other hand, the recovered metallicity distribution of the youngest burst (1--50 Myr) shows a significant spread around the input value. This uncertainty is due to the fact that this young episode is mostly populated by bright, massive stars in the upper-MS, for which the opacity is dominated by scattering with free electrons, and thus independent of $Z$. The normalised residuals confirm the difficulties in recovering the oldest episode of SF, showing a slightly systematic trend in the RGB region (see the slight excess of red colour around $(V-I) \sim 1$ and $25\lesssim I\lesssim26$). Nevertheless, all the other stellar evolutionary phases are well reproduced by the best model, with $90\%$ of the residuals within $\pm2 \sigma$.  

\end{document}